\documentclass[3p]{elsarticle}
\usepackage{epstopdf}
\usepackage{lineno,hyperref}
\usepackage{subfigure}
\usepackage[fleqn]{amsmath}
\modulolinenumbers[5]

\journal{Nuclear Instruments and Methods in Physics Research A}

\begin{document}

\begin{frontmatter}

\title{Systematical design and three-dimensional simulation of X-ray FEL oscillator for Shanghai Coherent Light Facility}

\author[mymainaddress,mysecondaryaddress]{Kai Li}

\author[mymainaddress]{Haixiao Deng
\corref{mycorrespondingauthor}}
\cortext[mycorrespondingauthor]{Corresponding author}
\ead{denghaixiao@sinap.ac.cn}

\address[mymainaddress]{Shanghai Institute of Applied Physics, Chinese Academy of Sciences, Shanghai 201800, China}
\address[mysecondaryaddress]{University of Chinese Academy of Sciences, Beijing 100049, China}

\begin{abstract}
Shanghai Coherent Light Facility (SCLF) is a quasi-CW hard X-ray free electron laser user facility which is recently proposed. Due to the high repetition rate, high quality electron beams, it is straightforward to consider an X-ray free electron laser oscillator (XFELO) operation for SCLF. The main processes for XFELO design, and parameters optimization of the undulator, X-ray cavity and electron beam are described. The first three-dimensional X-ray crystal Bragg diffraction code, named BRIGHT is built, which collaborates closely with GENESIS and OPC for numerical simulations of XFELO. The XFELO performances of SCLF is investigated and optimized by theoretical analysis and numerical simulation. 
\end{abstract}


\end{frontmatter}


\section{Introduction}

With the wide applications of X-ray in physics, chemistry and biology researches, facilities which is able to generate X-ray are under development in the last century continuously. Free electron laser (FEL) is a novel light source which produces high-brightness X-ray pulse \cite{deacon1977first,barletta2010free}. For pursuing high intensity and ultra-fast short wavelength radiation, some X-ray FEL facilities have been completed or under construction around the world\cite{altarelli2006european,ganter2010swissfel,emma2010first}.  Almost all of them take the advantages of self-amplified spontaneous emission (SASE) scheme \cite{bonifacio1984collective}, which starts from the electron beam shot noise and generates poorly temporal coherence light pulses. In order to obtain fully coherent X-ray pulses, Linac Coherent Light Source employs self-seeding method \cite{geloni2011novel,amann2012demonstration}, which is capable of improving the longitudinal coherence by a factor of 50. Nevertheless, the large pulse-to-pulse energy jitters (around 50\% r.m.s fluctuation) prevent it from further improvements.

An alternative promising approach to obtain fully coherent stable FEL pulses is X-ray free electron laser oscillator (XFELO) \cite{kim2008proposal,dai2012proposal}. Using the relativistic electron beam as gain medium to amplify the radiation trapped in an optical resonator is proposed conceptually and demonstrated experimentally decades ago in long wavelength region, such as for infrared and ultraviolet light \cite{billardon1983first,yan2016storage,oepts1995free}, but no XFELO experiment has been conducted due to the absence of X-ray high reflectivity mirrors. However, the X-ray high-reflectivity sapphire crystal experiment which was first demonstrated in 2010 \cite{shvyd2010high}, paved the way for real construction of XFELO. Taking the advantage of the crystal Bragg diffraction (BD), the XFELO scheme has been reconsidered and some practical technical problems has been studied \cite{song2016numerical,li2017simplified}.

\begin{figure}
   \centering
   \includegraphics*[width=220pt]{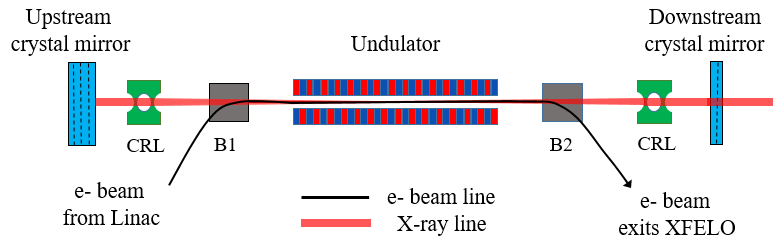}
   \caption{Schematic view of  XFELO for SCLF.}
   \label{schematic}
\end{figure}

Motivated by the successful operation of worldwide X-ray FEL facilities and the great breakthroughs in observation and control of very fast phenomena at the atomic time scale \cite{bostedt2016linac}, the first hard X-ray FEL light source in China named Shanghai Coherent Light Facility (SCLF) is under designing. SCLF plans to utilize photocathode electron gun combines with the superconducting Linac to produce 1 MHz repetition rate, 8 GeV relativistic electron beams. SCLF is capable of accommodating six undulator lines, and three of them will be constructed in the first phase project. After going through a beam switch yard, electron bunches are separated to three undulator lines which cover the photon energy of 0.4-25 keV. The undulator line that typically generates 12.4 keV X-ray, holds great potential for XFELO operation, as the layout shown in Fig.~\ref{schematic}. The cavity length is nearly 150 m, which ensures the X-ray pulse to meet an electron beam after being reflected back by the upstream crystal mirror at each round trip. The X-ray radiation overlaps with electron beam and extracts energy from it in the undulator. The upstream crystal is thicker than the downstream one, so that by transmission of the downstream mirror, part of X-ray pulse energy is coupled to the subsequent X-ray beam line. And the total reflectivity is assumed to be $R$. In order to maintain the growth of radiation power, the number of undulator periods should be large enough for single pass gain $G$ overcomes the round trip loss. According to \cite{boscolo1982gain}, the FEL gain is a constant at small signal regime, and decreases gradually due to the over modulation of electron beam when XFELO approaches saturation. The light pulse starts up from initial spontaneous radiation which acts as a seed and grows exponentially until reaches equilibrium, when the round trip energy net loss equals to the single pass gain. The light pulse power $P_n$ at the entrance of undulator for the $n$-th round trip is
\begin{equation}\label{power}
    P_n=P_{n-1} \times (1+G) \times R
\end{equation}

The main parameters of SCLF is presented in Table.~\ref{e-beam}. The bunch charge is tunable between 10 pC to 300 pC and the peak current is able to achieve several kA. Taking the advantages of superconducting accelerator technique, the quasi-CW electron bunches are suitable for XFELO operation. This facility, however, is basically designed for SASE and self-seeding, and the undulator periods has been optimized to be 26 mm with 5 m for each segment. This paper will discuss the XFELO operation on the condition of given electron beam and undulator parameters. In the next section, the processes and major concerns about XFELO design are discussed. And then the three-dimensional (3D) Bragg diffraction code BRIGHT is built for numerical simulations. The XFELO performances for the high peak current and low peak current mode are illustrated in Section~\ref{sec:level4} and Section~\ref{sec:level5}, respectively, 
and finally a brief summary is presented.

\begin{table}
   \centering
\caption{\label{e-beam}%
The main parameters of SCLF.
}
\begin{tabular}{lcr}
\hline
\textrm{Parameter}&
\textrm{Value}\\
\hline
Beam energy         & 8 GeV    \\ 
Relative energy spread     & 0.01 \%     \\ 
Repetition rate     & 1 MHz     \\
Peak current (low mode)      & 10, 20, 30 A    \\ 
Bunch charge (low mode)       & 20 pC     \\
Normalized emittance (low mode)       & 0.2 mm$\cdot$mrad     \\
Peak current (high mode)      & 0.5, 1, 1.5 kA    \\
Bunch charge (high mode)       & 100 pC     \\
Normalized emittance (high mode)       & 0.4 mm$\cdot$mrad     \\
Undulator period length    & 26 mm     \\
Undulator module length    & 5 m     \\
XFELO photon energy     & 14.3 keV     \\
\hline
\end{tabular}
\end{table}

\section{\label{sec:level2}XFELO design}
The process of XFELO design mainly contains two parts: the transverse and longitudinal profile optimization. On the one hand, electron beam transverse size is shaped by FODO (focusing-drift-defocusing-drift) lattice and optical transverse profile is determined by the cavity configurations. The goal is to improve coupling factor between electron bunch and radiation pulse in order to improve single pass gain. On the other hand, the longitudinal optimization concerns the required undulator length for sufficient gain and the optimum crystal mirror reflectivity for maximize the output pulse energy. For illustration, the 1 kA peak current beam and the corresponding parameters are utilized in the following subsections discussion.

\subsection{Undulator lattice matching}
For high energy electron beam, the transverse focusing from planar undulator are trivial \cite{quattromini2012focusing,qin2013design}, thus the external FODO structure is necessary. The quadrupole magnets are installed at the nearly 1 m drift space between two undulator modules and the Twiss parameters are matched to let the beta function evolves periodically. For a given bunch charge, the smaller transverse beam size is, the higher peak current as well as single pass gain will be \cite{krinsky1987output,kim1992fel,chin1993calculation}. In addition, small beam cross section is also good for preserving the X-ray fundamental transverse mode \cite{moore1985high}.

\begin{figure}[!htb]
   \centering
   \includegraphics*[width=220pt]{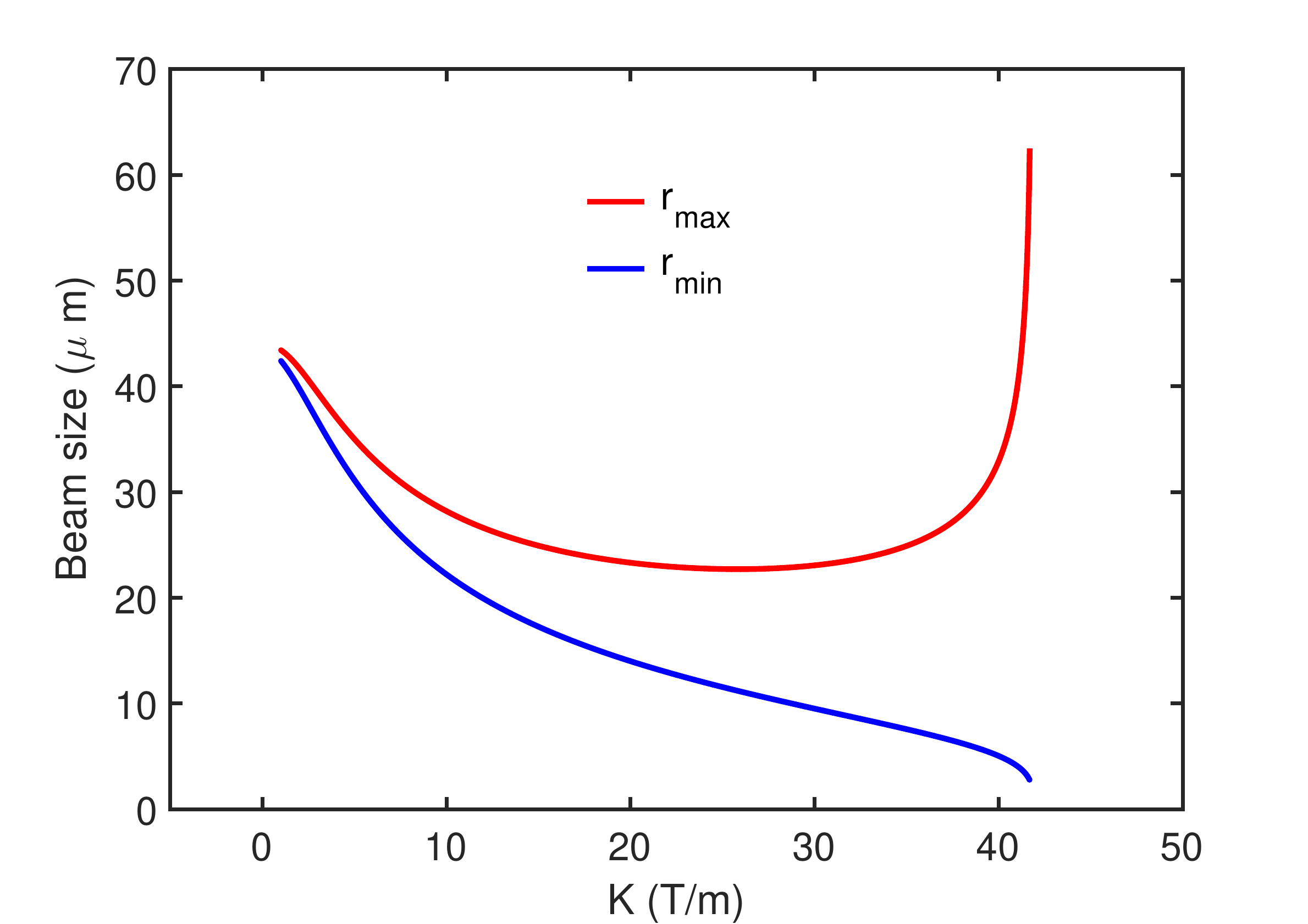}
   \caption{Transverse electron beam size inside the FODO lattice for different quadrupole focusing strength. The upper (red) and the lower (blue) line show the maximum and minimum beam size along the undulators, respectively.}
   \label{beamsize}
\end{figure}

The lattice is matched by matrix method \cite{wiedemann2015particle}. The electron beam transverse beta function varies at different position of FODO, and its maximum and minimum radius $r$ change with different quadrupole intensity are shown in Fig.~\ref{beamsize}. The drift space between two modules and the quadrupole length is assumed to be 1 m and 20.8 cm, respectively. When focusing strength is small, the electron beam size expands and exhibits little fluctuation; as the quadrupole intensity increases, the average beam size declines gradually, however, the corresponding beam radius oscillation increases; and when the focusing strength is too strong, no periodical solutions exist for the beam dynamic functions, so the beam size explodes rapidly. For pursuing maximum gain, we chose the quadrupole magnet gradient to be 20 T/m to get the small average beam size with relatively low variation between 15 $\mu$m and 25 $\mu$m.

\subsection{Optical cavity design}
\begin{figure}[!htb]
   \centering
   \includegraphics*[width=220pt]{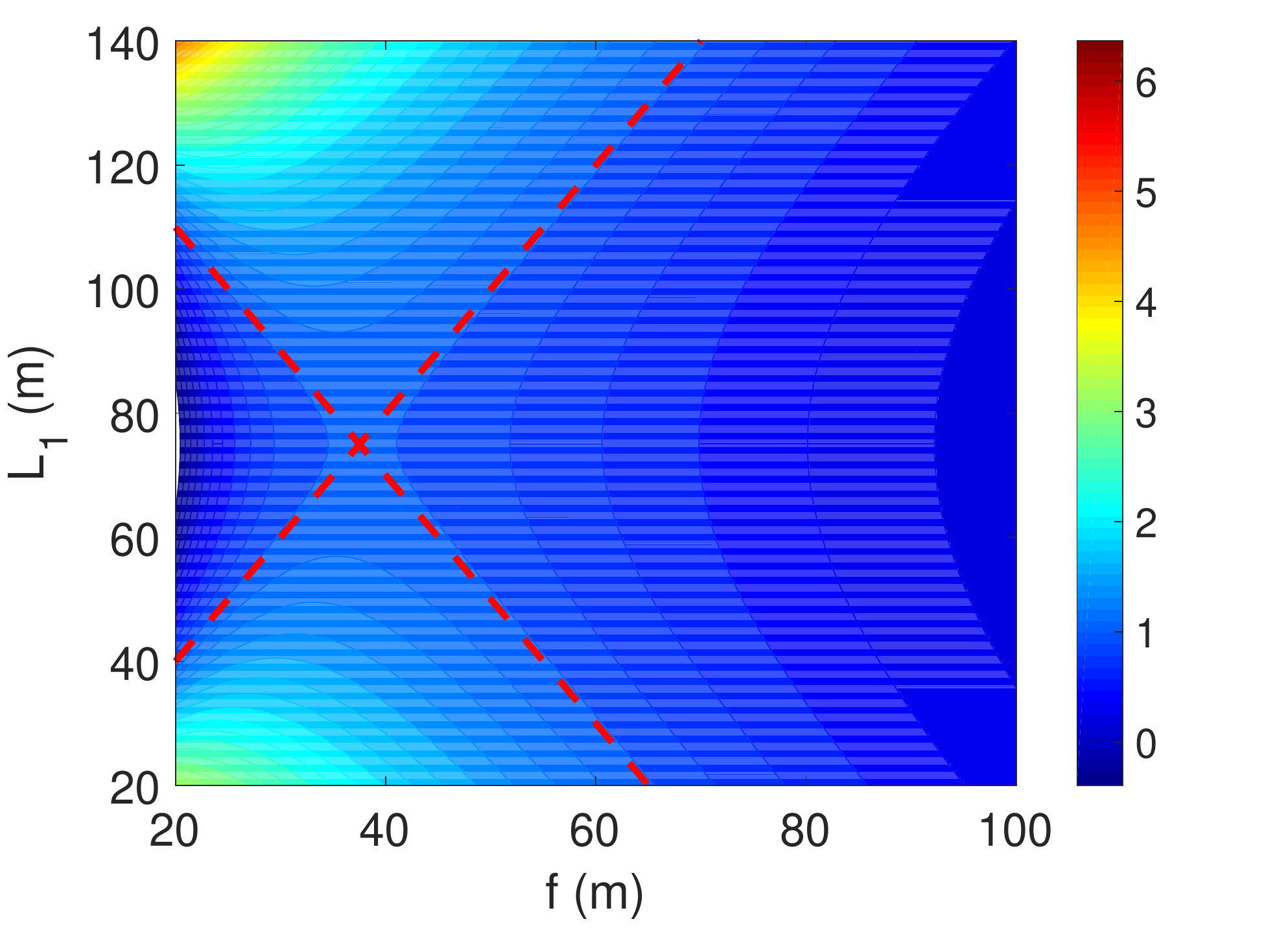}
   \caption{The value of parameter $|m|$ as a function of CRLs focal length and the distance between them.}
   \label{cavity}
\end{figure}
As aforementioned, the $L=150$ m symmetry resonator is formed by two sapphire crystal mirrors. According to \cite{lindberg2011performance}, X-ray focusing elements are required for the stability of optical cavity, thus two Be parabolic compound refractive lenses (CRLs) \cite{lengeler1999imaging} with focal length $f=57.7$ m are used, each with a radius of 0.33 mm and a very high transmissivity $T=0.987$, assuming surface microroughness less than 0.3 $\mu$m and an effective aperture of $0.66$ mm. The rays propagation inside cavity is analyzed by ABCD matrix methods \cite{siegman1986lasers}, which defines the ray matrix for propagation through one period of the optical system as
\begin{equation}\label{matrix}
    M\equiv\begin{bmatrix} A & B \\ C & D \end{bmatrix}; \qquad m\equiv\frac{A+D}{2}
\end{equation}
For the periodic focusing system to be stable, the eigenvalues of $M$ should be complex and have magnitude unity thus $|m|<1$. Figure.~\ref{cavity} presents the contour plot of $|m|$ parameter as a function of CRLs focal distance $f$ and the space $L_1$ between them. The red dashed line denotes the points where $m=-1$, and the blue regions inside two critical lines are favorable values for stable cavity configuration. According to it the distance between two CRLs is 34.6 m$<L_1<115.4$ m. Note that different from the high-gain FEL in which transverse radiation distribution is determined by gain guiding effect. The XFELO optical eigenmodes depend on the configurations of resonator. For the coupling benefit between electron bunch and X-ray pulse, $L_1$ is optimized to be 100 m to match electron beam average radius by optical propagation code (OPC) \cite{van2009time}.

\begin{figure}[!htb]
   \centering
   \includegraphics*[width=220pt]{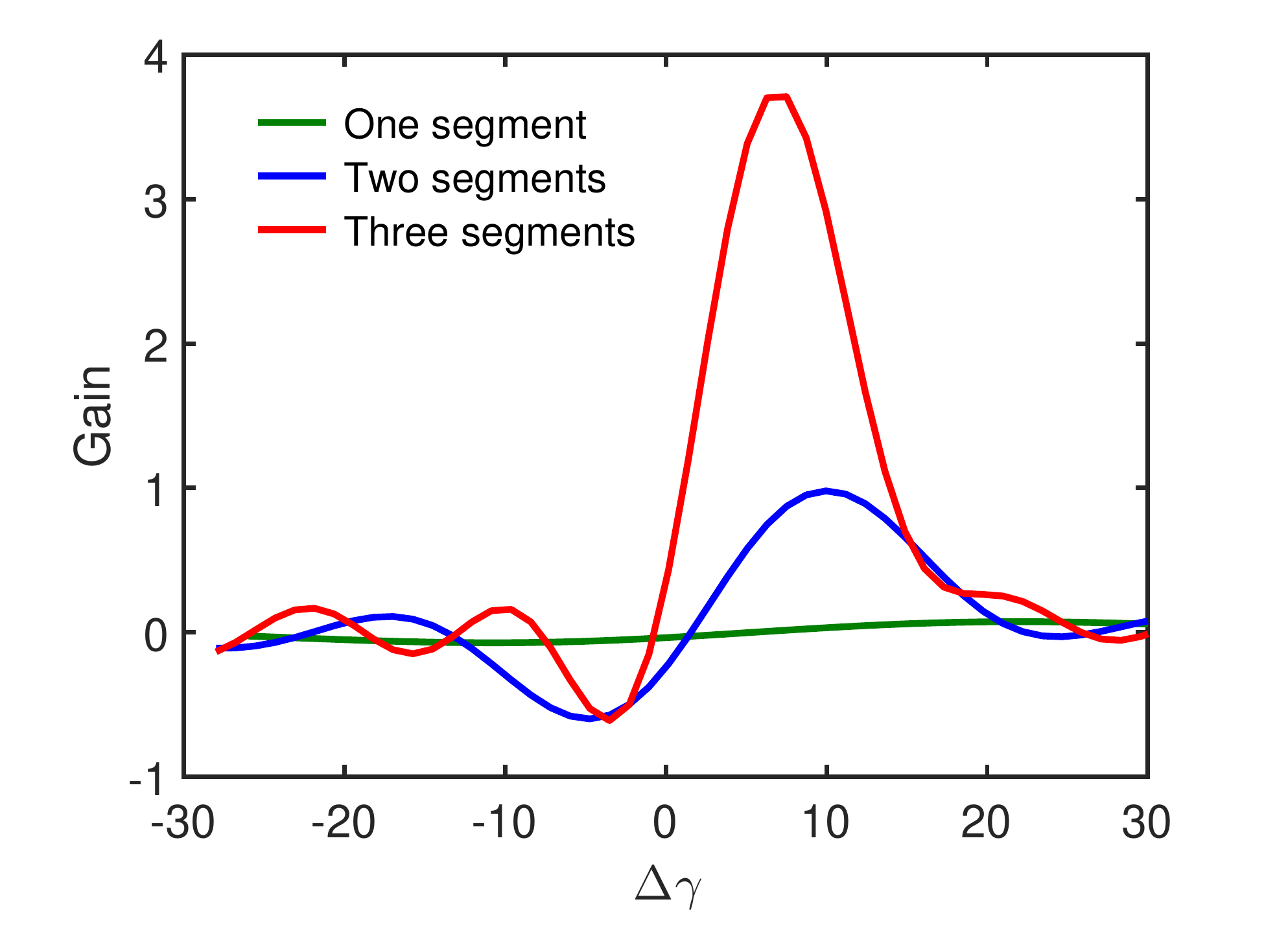}
   \caption{Single pass gain curve with different undulator segments.}
   \label{gain}
\end{figure}

\subsection{Undulator length}
For FEL oscillator, the undulator length is a trade-off between high power and sufficient single pass gain. On the one hand, the maximum energy extracted from a FEL oscillator is inversely proportional to the number of undulator period \cite{kroll1981free}. On the other hand, the undulator should be sufficient long to ensure single pass gain overcomes round trip loss. Thus the undulator is chosen to be as short as possible once providing enough gain. Fig.~\ref{gain} demonstrates the single pass gain as a function of  $\Delta\gamma$, which is the deviation of scaled electron beam energy from resonant energy. The curve displays a typical small signal gain function, and for different numbers of undulator periods, the maximum gain achieved at different magnitude of energy deviation, and it approaches resonate energy as the undulator length enhances. For a given beam energy, the undulator gap is adjusted to obtain proper energy detune in practice. Considering the cavity passive loss ($\sim$10\%), the sapphire crystal finite bandwidth acceptance loss plus absorption ($\sim$40\%) and cavity output coupling ($\sim$20\%), the optimized undulator length is three segments.

\subsection{Cavity crystal mirror}
\begin{figure}[!htb]
   \centering
   \includegraphics*[width=220pt]{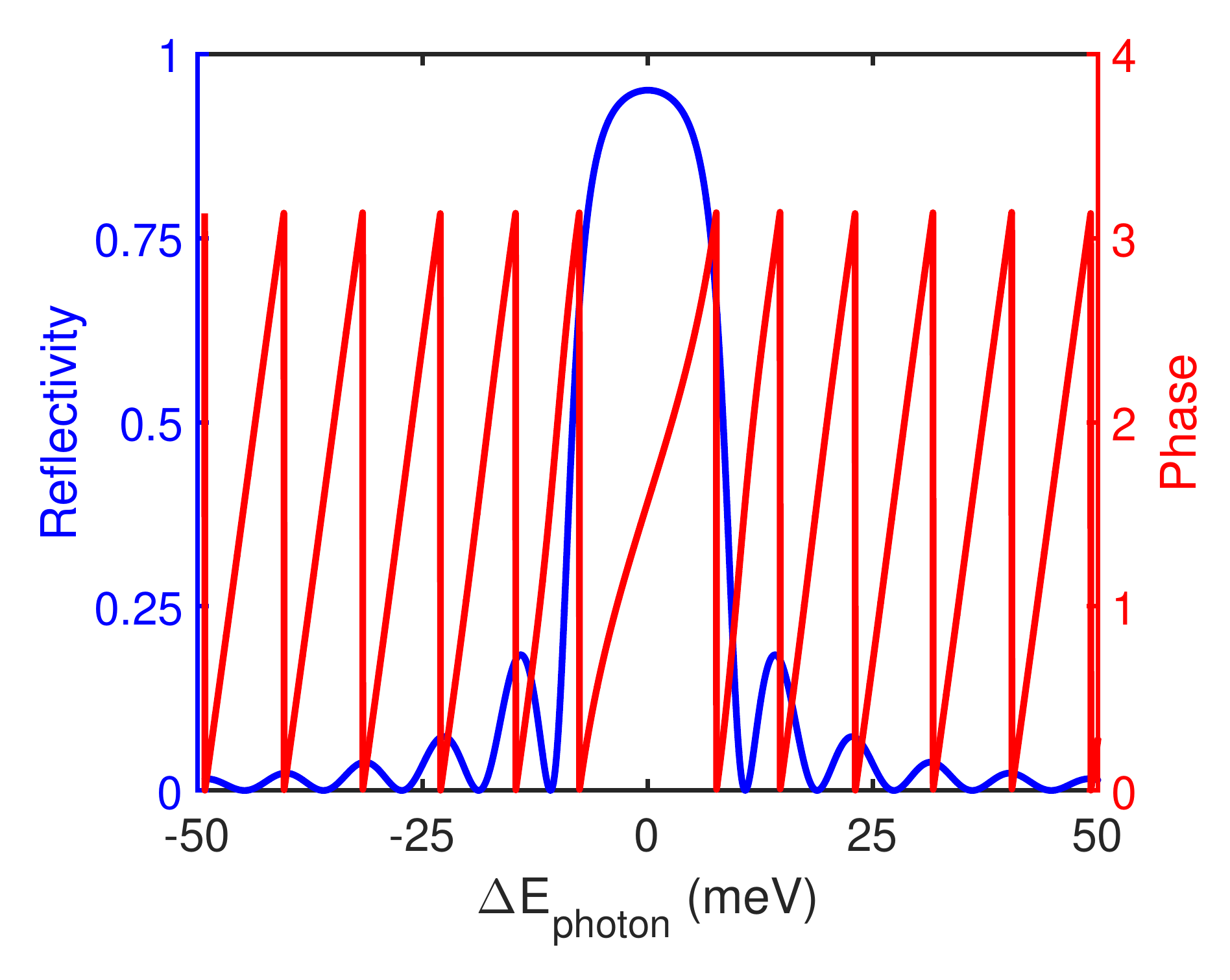}
   \caption{The crystal mirror complex reflectivity for different incident photon energy.}
   \label{Bragg_reflectivity}
\end{figure}

XFELO employs crystal Bragg diffraction, which enables high reflectivity at a narrow bandwidth. According to \cite{shvyd2012spatiotemporal}, the response function is depends on crystal thickness, X-ray incident angle, and excitation length. Assuming that the interaction is Bragg symmetry diffraction, and the thickness of crystal is designed for different reflectivity. For pursuing high reflectivity at normal-incidence, low photon-electron effects and high Debye temperature, such as sapphire crystal with (0 0 0 30) atom plane is used for 14.3 keV photon Bragg exact backscattering. Figure.~\ref{Bragg_reflectivity} shows the theoretical reflectivity for 70 $\mu$m thickness sapphire crystal with normal incidence. A high reflectivity nearly 95\% within 13 meV Darwin width was predicted. Note that the phase has been manually scaled into $(0, 2\pi)$, the phase deviation due to Bragg diffraction insides Darwin width produces an additional delay to X-ray pulse, which should be compensated by cavity detune to avoid FEL gain degradation. 
\begin{figure}[!htb]
   \centering
   \includegraphics*[width=220pt]{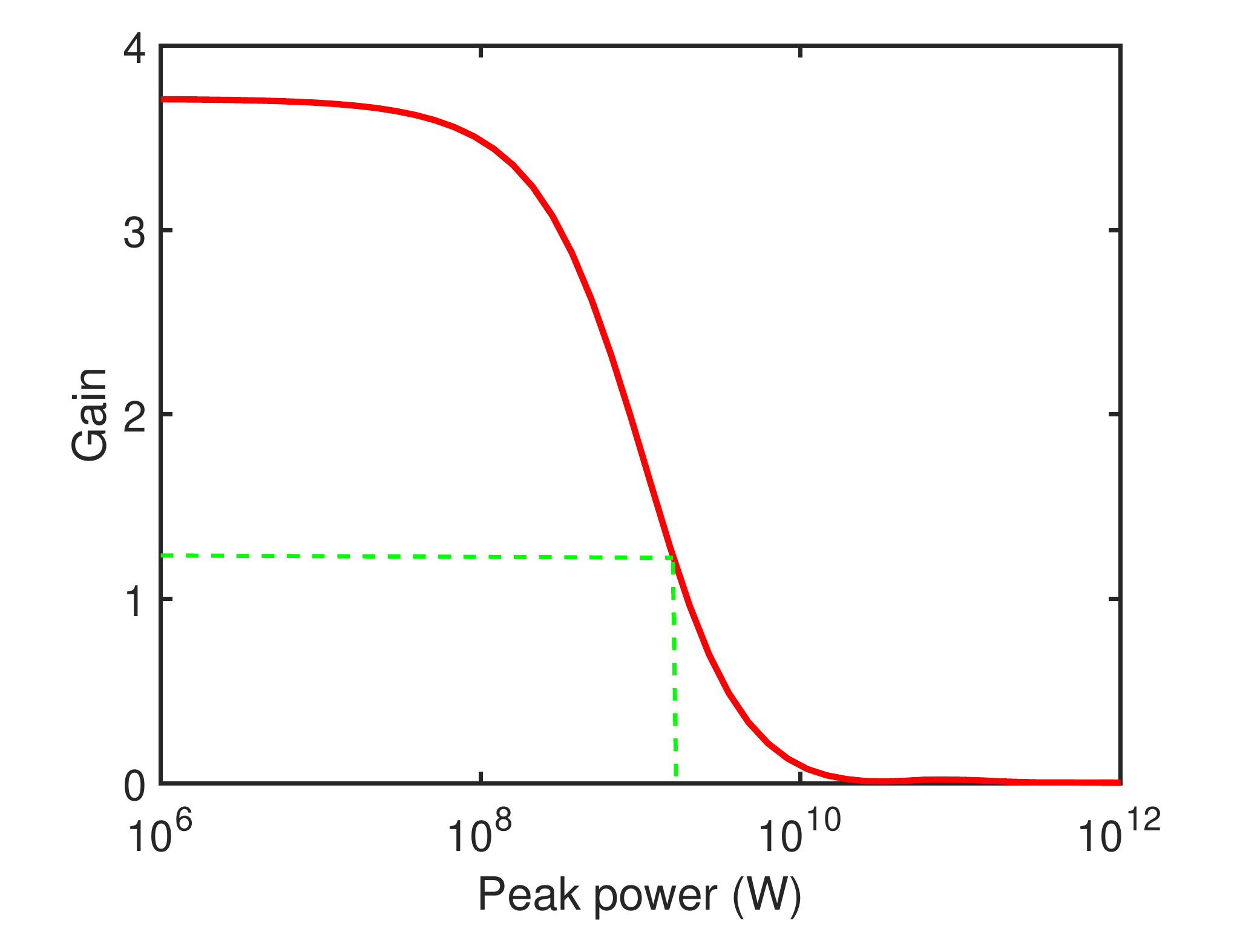}
   \caption{The gain various at different radiation power. The green line shows a typical XFELO round trip loss nearly 57\% (including 20\% output coupling), and the saturation peak power inside the cavity would be around 1.5 GW, thus the output power is 0.3 GW.}
   \label{gain_power}
\end{figure}

The cavity output coupling is optimized to generate highest X-ray pulse energy. Fig.~\ref{gain_power} shows a typical time-independent XFELO gain as a function of light pulse peak power inside the cavity. When radiation intensity is small, the single pass gain remains constant. While X-ray reaches saturation, portion of electrons undergo strongly FEL interaction beyond 1/2 synchrotron oscillation and energy extraction efficiency drops gradually. When single pass gain equals to the round trip cavity net loss, the net gain is zero, and XFELO reaches equilibrium and the output pulse energy remains steady.

\begin{figure}[!htb]
   \centering
   \includegraphics*[width=220pt]{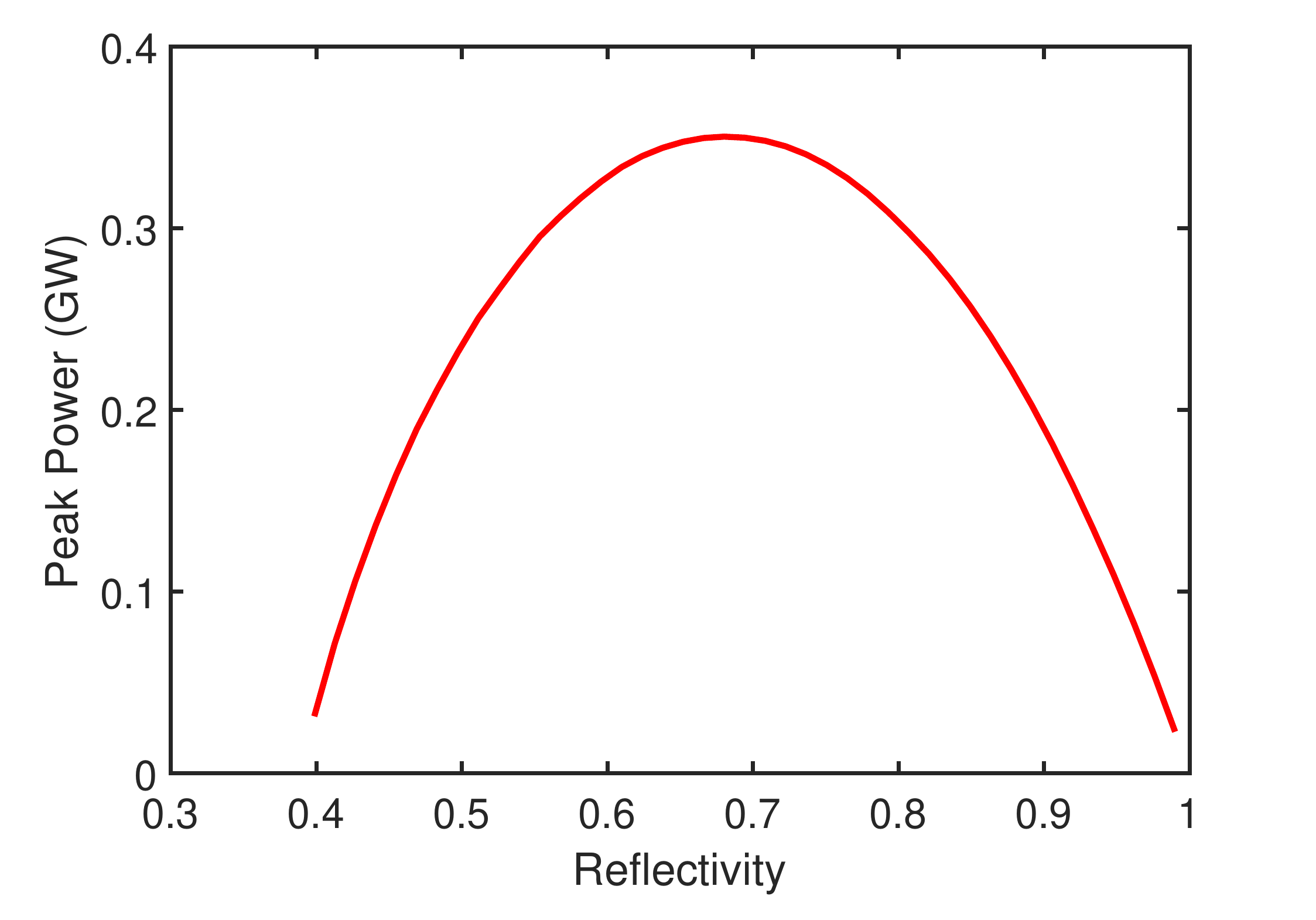}
   \caption{The output peak power at different mirror reflectivity.}
   \label{loss}
\end{figure}

The peak power inside the cavity is calculated by interpolation, like in Fig.~\ref{gain_power}, and the output power is predicted by multiplying mirror transmissivity. Fig.~\ref{loss} represents the output X-ray peak power as a function of mirror reflectivity, in which the cavity round trip net passive loss, including diffraction loss and CRLs absorption, is assumed to be 10\%. On the one hand, for the low reflectivity cavity, due to large amount of pulse energy leaks out the optical cavity, effective positive feedback cannot build up thus small radiation peak power; On the other hand, when the reflectivity is too high to let the light power coupling out efficiently, the X-ray output power is small as well. The optimum reflectivity is nearly 70\% which predicts around 0.35 GW output peak power.

\section{\label{sec:level3}Three-dimensional effect of Bragg diffraction}
\begin{figure*}
   \centering
   \includegraphics*[width=380pt]{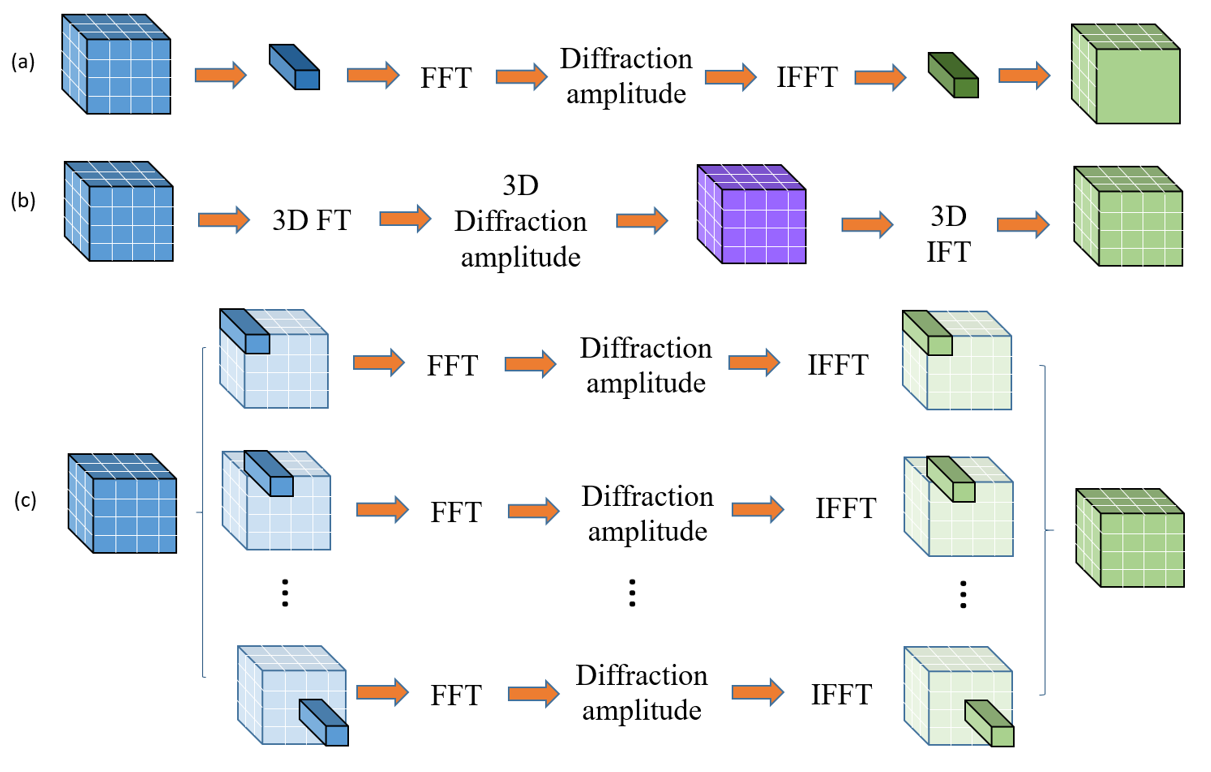}
   \caption{The schematic of calculating Bragg diffraction. (a) Traditional one dimensional process, which losses the transverse phase information; (b) The directly three-dimensional way, which preserve the transverse phase distribution and is supposed to be precise, while is much more time-consuming; (c) The new approach which takes advantages of Fast Fourier Transform and enjoys high efficiency and accuracy.}
   \label{bragg}
\end{figure*}

The FEL oscillator simulations are normally carried out by combination of 3D FEL codes GENESIS \cite{reiche1999genesis} and OPC. As demonstrated in \cite{karssenberg2006modeling,jin2012numerical,tin2016design}, these codes are efficient for long wavelength FEL oscillator simulations like ultraviolet or infrared light. However, for XFELO which includes the interaction between X-ray pulse and crystal, some extensions to the classical methods are needed. In the past, as far as we know, the Bragg diffraction is traditionally calculated in one or two dimensional way \cite{shvyd2012spatiotemporal,yang2013maximizing}, i.e., the 3D radiation field is usually transformed into one dimension for crystal reflection and then artificially transformed back to 3D in simulation, as seen in Fig.~\ref{bragg}(a). It smears out the transverse phase information totally and ignores some important 3D effects including wave-front, which is essential for transverse modes and cavity stability. The defects are not significant when cavity configuration is primitive or single pass gain is overwhelming, but for more advanced and delicate resonators \cite{lindberg2011performance}, the self-consistent XFELO simulations including all 3D effects are preferred \cite{geloni2011novel}.

According to \cite{shvyd2012spatiotemporal}, Bragg diffraction of X-ray pulse was studied by taking Fourier transform (FT), which decomposes the light pulse into monochromatic plane-wave components, and multiply the corresponding Bragg diffraction amplitude with each of them. This approach is available for 3D simulation: the spatiotemporal radiation distribution function undergoes Fourier transformation, then the related Bragg reflection spectral amplitude $R_{0H}$ are multiplied by each monochromatic planar wave, finally inverse Fourier transform (IFT) is unitized to obtain the diffraction radiation. 

Following the notation and deduction of two-dimensional spatiotemporally response of ultrashort, laterally confined X-ray pulses Bragg diffraction in \cite{shvyd2012spatiotemporal,shvyd2013x}, we present here the derivation of a finite long X-ray pulse three-dimensional Bragg diffraction, which provides theoretical support of a new calculation method. The response for a monochromatic planar X-ray wave $\varepsilon_i$ with a frequency $\omega_0$, incident angle $\theta$ respect to atom plane is
\begin{equation}
\varepsilon_{H}^{(m)}(t,\vec{r})=\varepsilon_i e^{ -i \left[ \omega_0 t - (\vec{K}_0+\vec{H})\vec{r} \right]} e^{i \left[\Delta_H (\vec{r}-\vec{r}_H)\hat{z} \right]} R_{0H}(\omega_0)
\label{a1}
\end{equation}
where
\begin{equation}
\vec{K}_0\approx\frac{\omega_0}{c}\hat{u}_0+\frac{\omega_0\, \delta\phi}{c}\hat{w}_0
\end{equation}
as defined in \cite{shvyd2012spatiotemporal}, $\hat{u}_0$ is the direction of optics propagating axis, while $\hat{w}_0$ is the direction perpendicular to the dispersion plane. The azimuthal angle $\delta\phi$, between $\vec{K}_0$ and dispersion plane, is presumed to be small for a typical low divergence X-ray FEL. Thus the wave number component in $\hat{w}_0$ is much smaller than along $\hat{u}_0$, the influence of it on the additional momentum transfer $\Delta_H$ is negligible. We eliminate all the higher order terms and follow the methods used in \cite{shvyd2012spatiotemporal}, rewrite Eq.~(\ref{a1}) as
\begin{equation}
\varepsilon_{H}^{(m)}(t,\vec{r})=\varepsilon_i \, exp\left( i\, \omega \,  \frac{\hat{w}_0 \cdot \vec{r}}{c} \delta \phi \right)\,exp\left(-i\,\omega \tau_H \right) exp\left(-i \, \Omega \, \xi_H\right) R_{0H}(\Omega+\omega)
\end{equation}
where
\begin{equation}
\tau_H=t-\frac{\hat{u}_H \cdot \vec{r}}{c}, \quad \xi_H=t-\frac{\hat{u}_0 \cdot \vec{r}}{c}-2 sin^2\theta\frac{(\vec{r}-\vec{r}_H)\hat{z}}{c \gamma_H}, \quad \omega+\Omega=\omega_0.
\end{equation}
and we have assumed that $\left|\Omega/\omega\right| \ll 1$ and omitted $\Omega\, \delta \phi$ term. Here $\omega$ is the Bragg frequency, $\omega_0$ is the frequency of incident photon.

For a finite long X-ray pulse with confined wave front, the radiation after Bragg diffraction would be
\begin{eqnarray}
\varepsilon_{H}(t,\vec{r})&=&\varepsilon_i \int_{-\infty}^{+\infty} \frac{d\tilde{\theta}}{2\pi}exp\left[- i \, \omega (\tilde{\theta})\tau_H(\tilde{\theta}) \right] \int_{-\infty}^{+\infty} \frac{d\,\delta \phi}{2\pi}exp\left[ i \,\omega (\tilde{\theta})\frac{\hat{w}_H \cdot \vec{r}}{c} \delta \phi \right] \nonumber \\
  & & \times \int_{-\infty}^{+\infty} \frac{d\Omega}{2\pi} exp\left(-i \, \Omega \, \xi_H\right) f(\Omega,\tilde{\theta},\delta \phi) R_{0H}(\Omega+\omega)
\label{3d_direct}
\end{eqnarray}
where $f(\Omega,\tilde{\theta},\delta \phi)$ is the Fourier transform of incident pulse, i.e.
\begin{eqnarray}
\varepsilon_{i}(\tau_0,\nu_0,w_0)&=&\varepsilon_i \,exp(-i\,\omega \tau_0) \int_{-\infty}^{+\infty} \frac{d\tilde{\theta}}{2\pi}exp\left[ i \, \nu_0 (\omega_0/c)(\tilde{\theta}-\theta) \right] \int_{-\infty}^{+\infty} \frac{d\,\delta \phi}{2\pi}exp\left[ i \, w_0 (\omega_0/c)\delta \phi \right] \nonumber\\
  & & \times \int_{-\infty}^{+\infty} \frac{d\Omega}{2\pi} exp\left(-i \, \Omega \, \tau_0\right) f(\Omega,\tilde{\theta},\delta \phi)\nonumber\\
  & \equiv & \varepsilon_i \,exp(-i\,\omega \tau_0) \int_{-\infty}^{+\infty} \frac{d\Omega}{2\pi} exp\left(-i \, \Omega \, \tau_0\right) \prod(\Omega,\nu_0,w_0)
\end{eqnarray}

This formula could be used for 3D Bragg diffraction simulation directly as shown in Fig.~\ref{bragg}(b), which is expected to be accuracy theoretically. In practice, however, it is proved to be time-consuming because the odd number of grid points required by GENESIS is not consistent with fast Fourier transform (FFT). With the absence of FFT, scheme (b) takes almost a whole week on a personal computer to obtain the results after Bragg diffraction, which is definitely not suitable for XFELO simulations. To simplify the Eq.~(\ref{3d_direct}), we may introduce some reasonable approximations \cite{shvyd2012spatiotemporal}:
\begin{align}
\tau_H (\tilde{\theta}) &\approx \tau_H + \frac{\nu_H}{c}(\tilde{\theta}-\theta), \qquad
\omega (\tilde{\theta}) \approx \omega[1-(\tilde{\theta}-\theta)\,cot\theta],\nonumber\\
\tau_H (\tilde{\theta})\,\omega (\tilde{\theta}) &\approx \omega \tau_H + \frac{\omega}{c}(\nu_H-\tau_H\,c\,cot\theta)(\tilde{\theta}-\theta), \qquad
\omega (\tilde{\theta}) \, \delta \phi \approx \omega \delta\phi
\end{align}
The spatiotemporal radiation after Bragg diffraction is rewritten as
\begin{equation}
\varepsilon_{H}(t,\vec{r}) \approx \varepsilon_i e^{-i\,\omega\tau_H}\int_{-\infty}^{+\infty} \frac{d\Omega}{2 \pi} e^{-i\,\Omega \xi_H} \prod(\Omega,\,\nu_H-\tau_H\,c\,cot\theta,\,w_H )R_{0H}(\Omega+\omega)
\end{equation}
where $w_H=\hat{w}_H\cdot\vec{r}$.

The results indicate that for an incoming high brightness X-ray pulse from FEL, as shown in Fig.~\ref{bragg}(c), the radiation after Bragg diffraction could be obtained by: first, Fourier transform each longitudinal wavelet at a fixed transverse coordinate; and then multiply the corresponding monochromatic wave diffraction amplitude function $R_{0H}$, and get the radiation wavelet after Bragg diffraction by inverse Fourier transform; finally, substitute the transverse variable $\nu_H$ by $\nu_H-\tau_H\,c\,cot\theta$ which represents spatial transverse shift. In practice, this could be taken into account by a linear coordinates transform, since only first order small terms are important and concerned in the derivation. For the XFELO discussed in this paper, the incident angle $\theta$ is approximate to $\pi/2$, which means $cot\theta \approx 0$ and there is not transverse deformation. According to the strategies (b) and (c) mentioned above, the BRagg dIffraction alGoritHom in Three-dimensional (BRIGHT) which is tailored for XFELO simulations has been built. Due to the benefits of FFT in scheme (c), only several seconds is required for obtaining the 3D spatiotemporal radiation after Bragg diffraction.

\begin{figure*}
   \centering
   \subfigure{\includegraphics*[width=100pt]{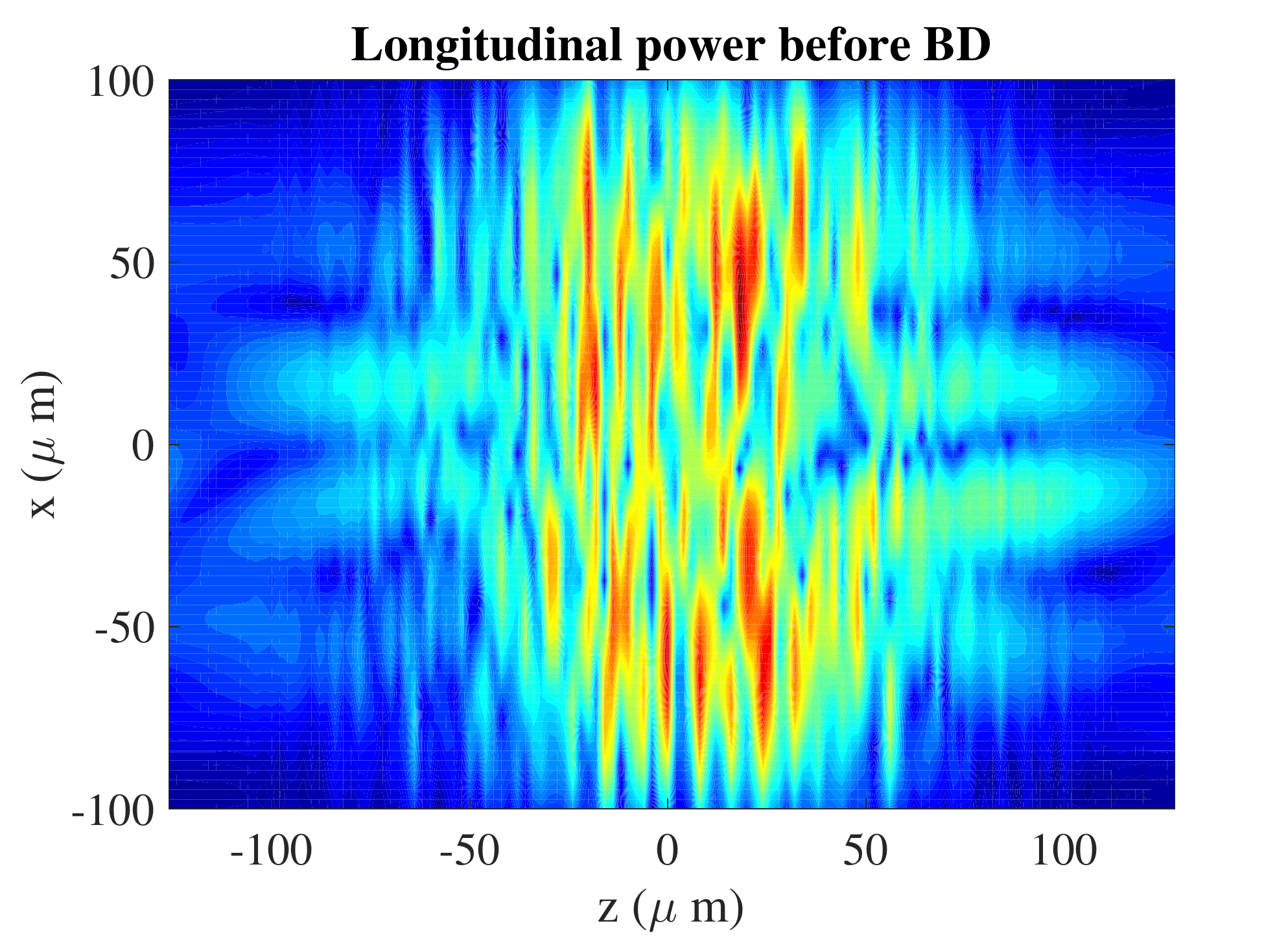}}
   \subfigure{\includegraphics*[width=100pt]{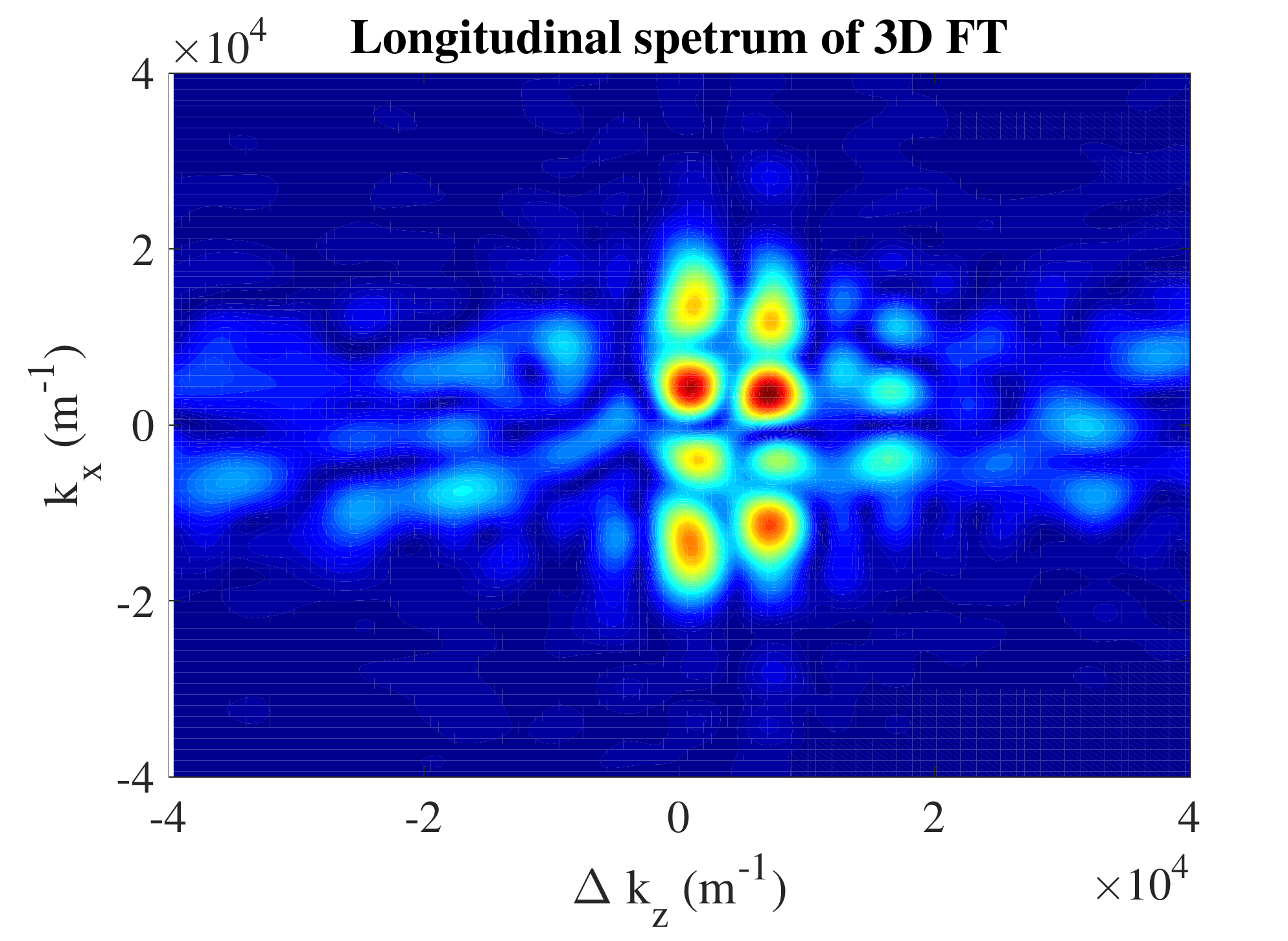}}
   \subfigure{\includegraphics*[width=100pt]{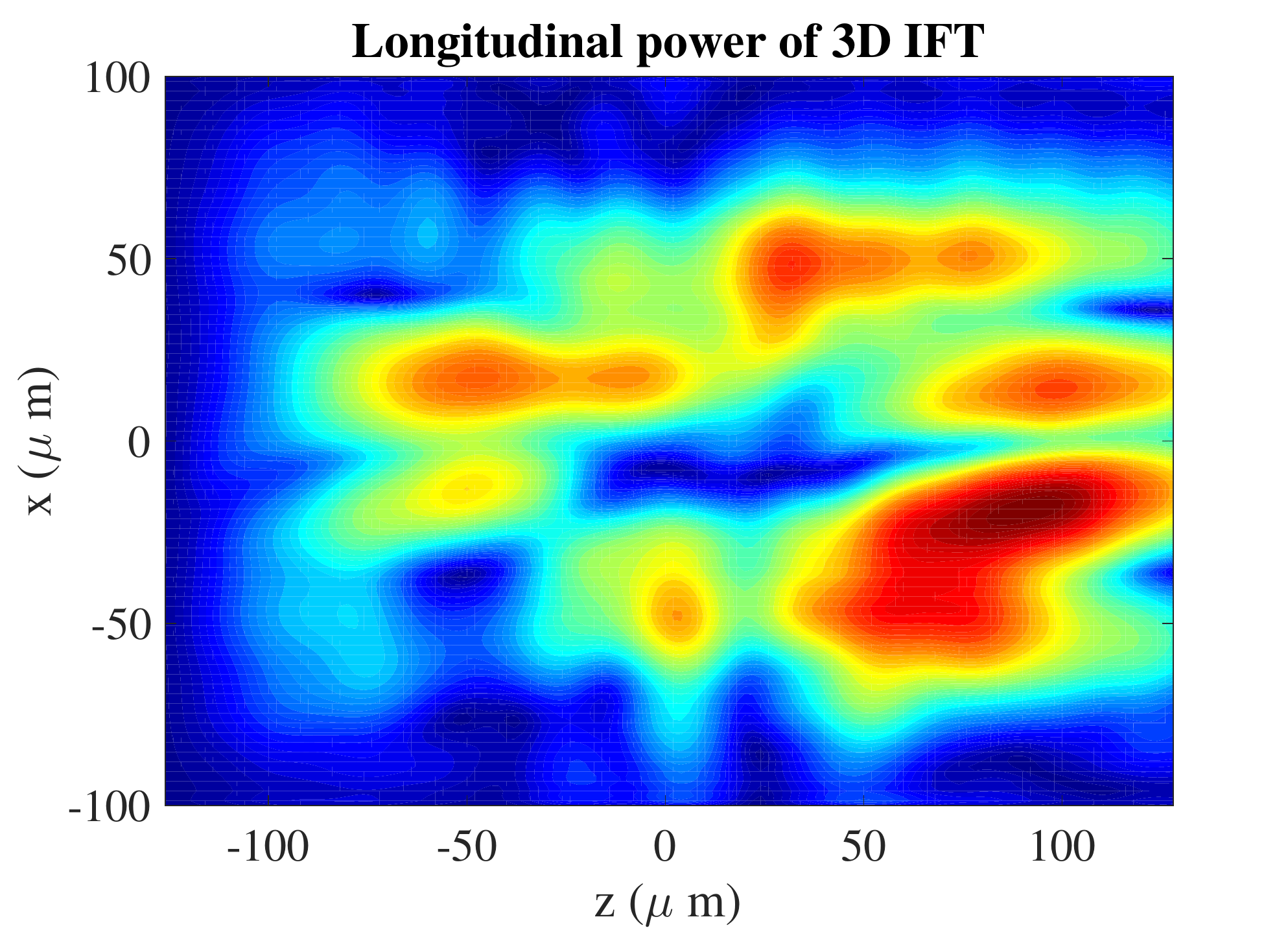}}
   \subfigure{\includegraphics*[width=100pt]{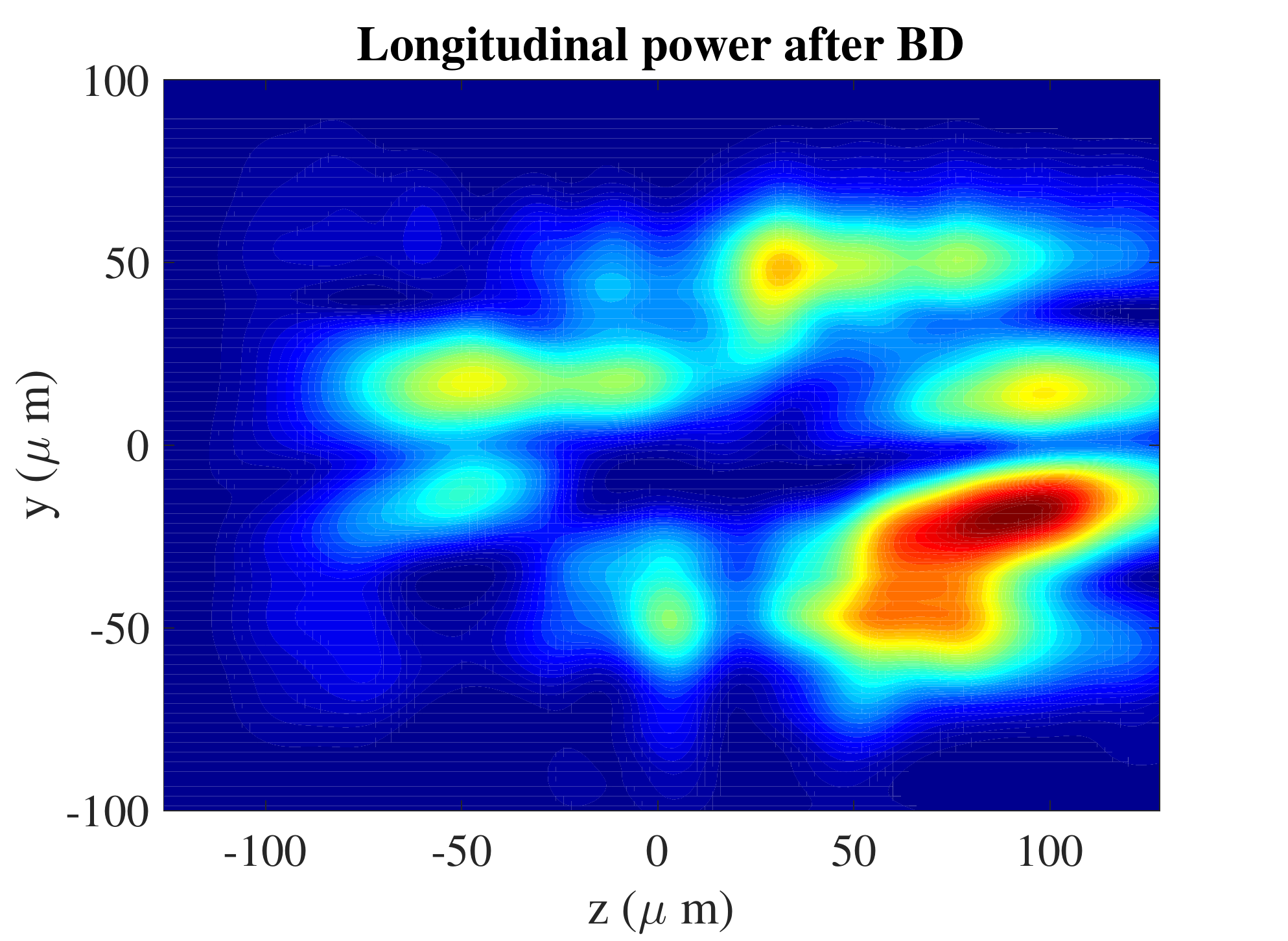}}
   \subfigure{\includegraphics*[width=100pt]{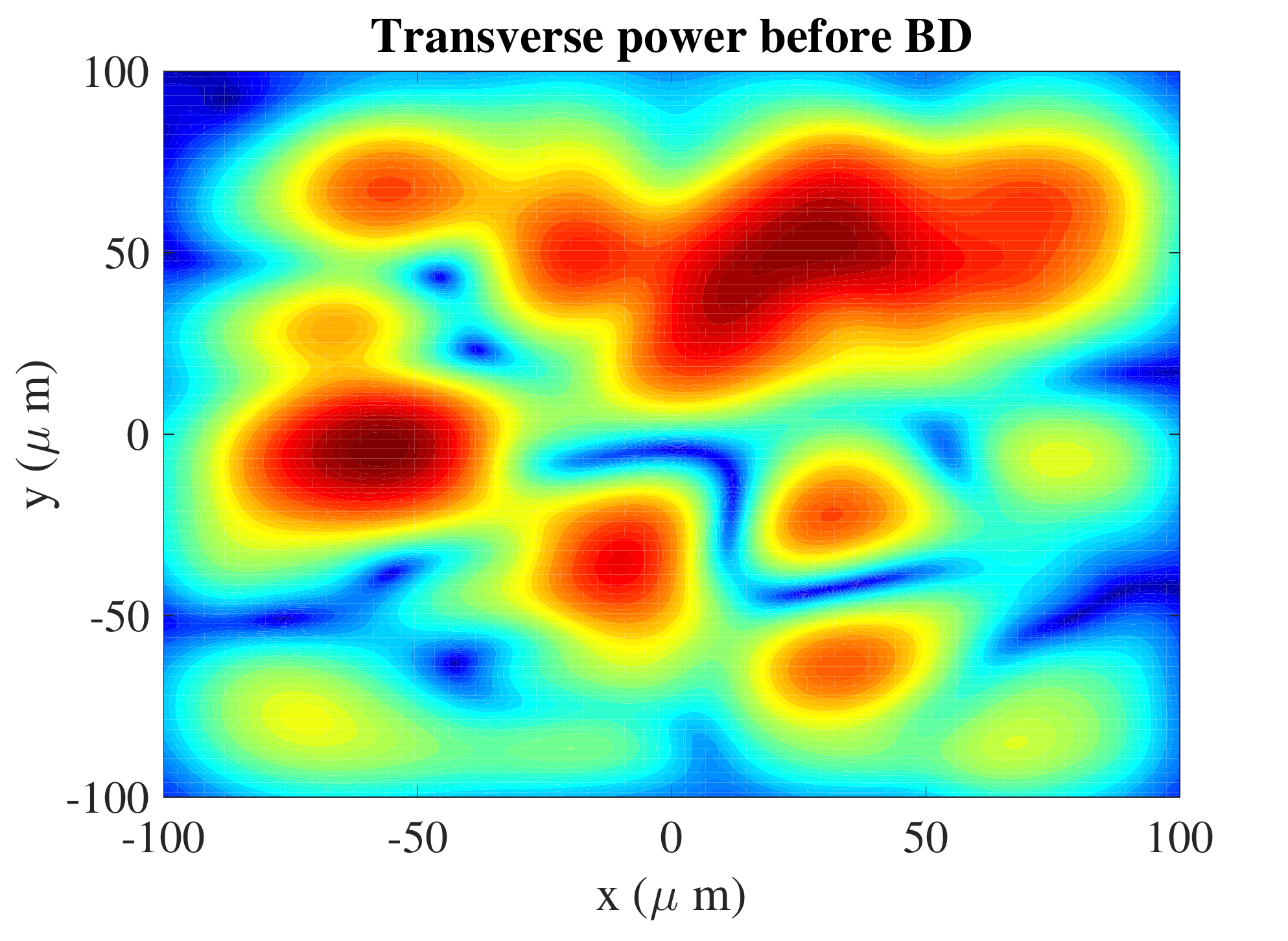}}
   \subfigure{\includegraphics*[width=100pt]{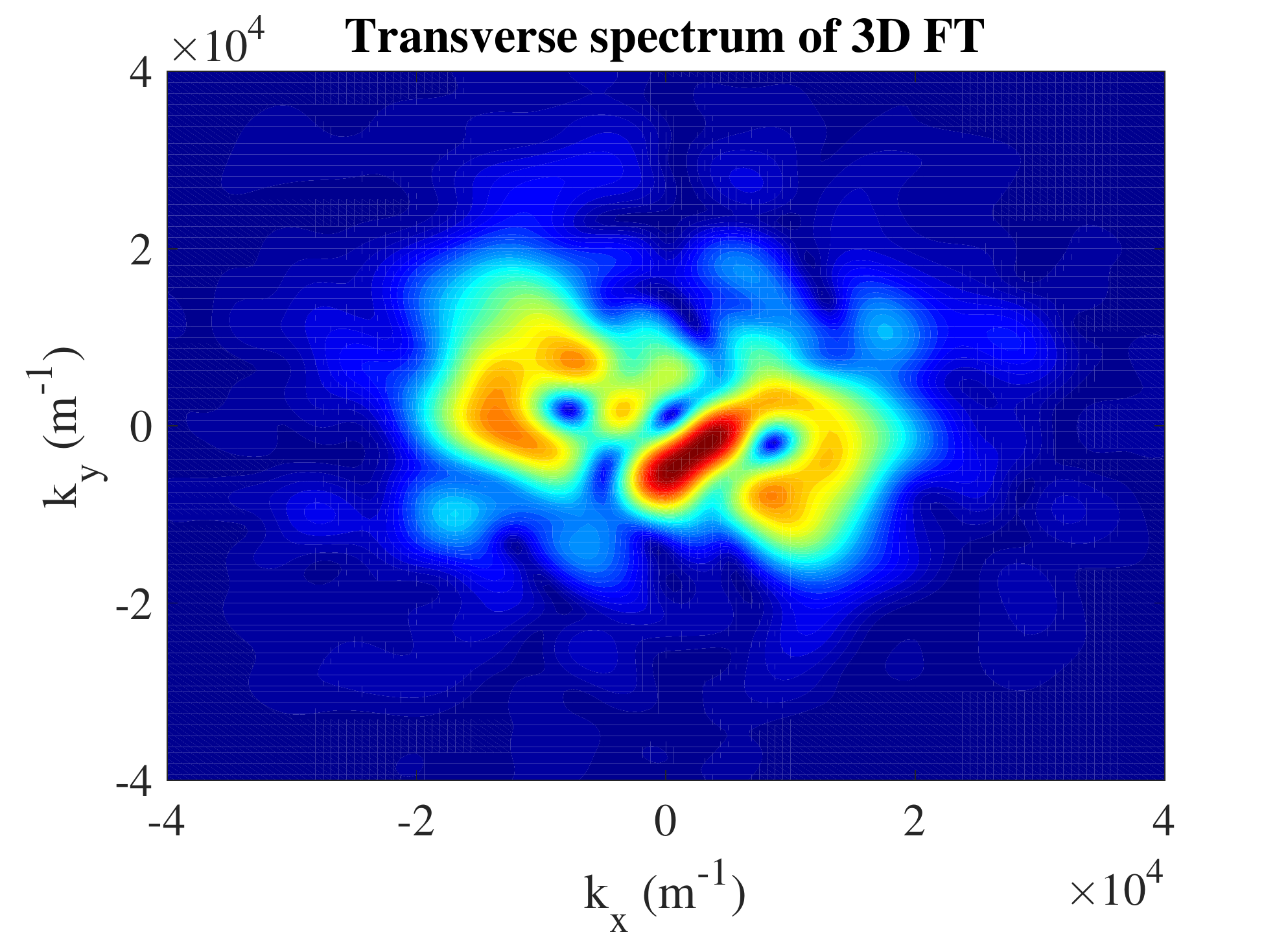}}
   \subfigure{\includegraphics*[width=100pt]{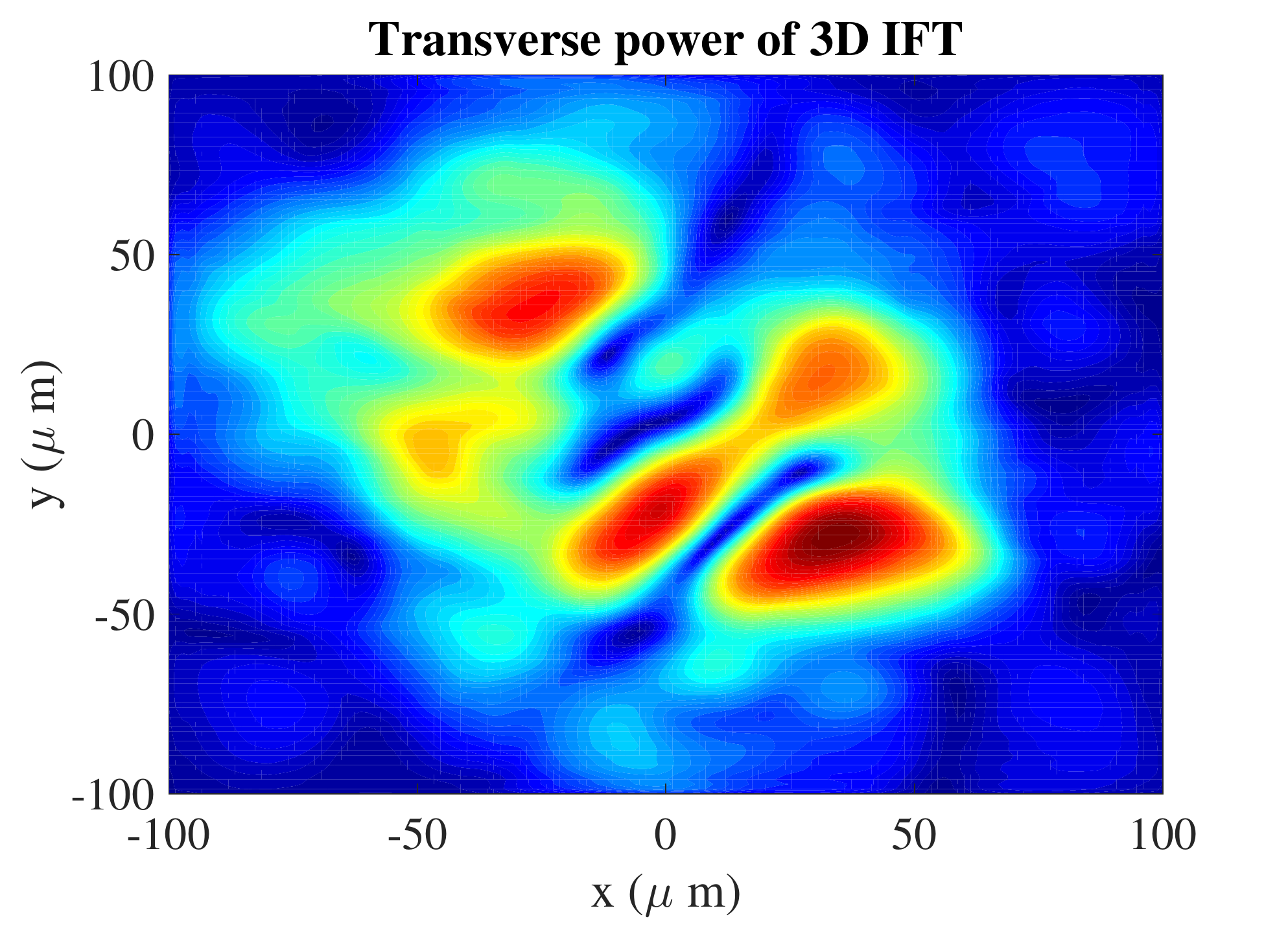}}
   \subfigure{\includegraphics*[width=100pt]{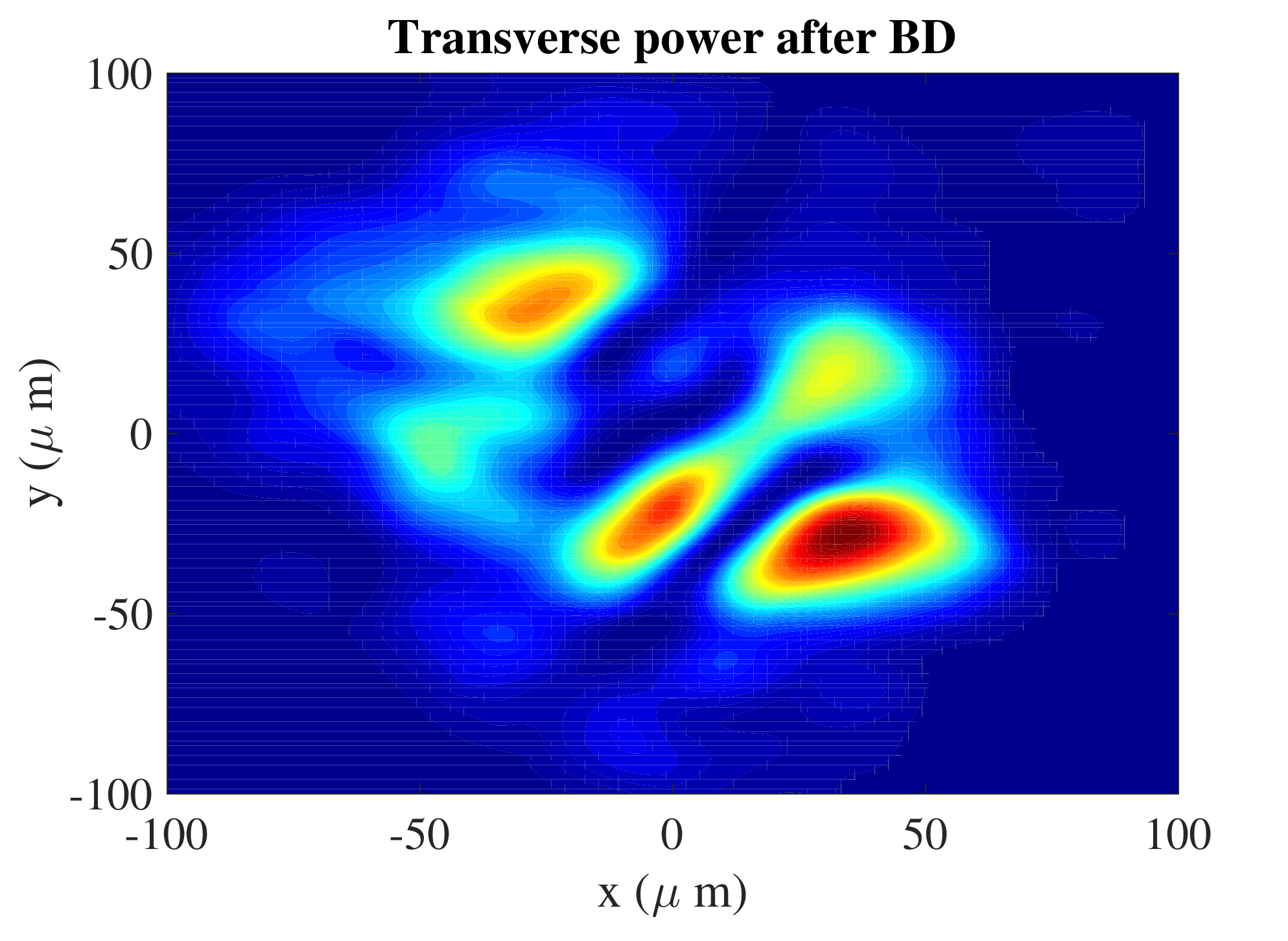}}
   \subfigure{\includegraphics*[width=100pt]{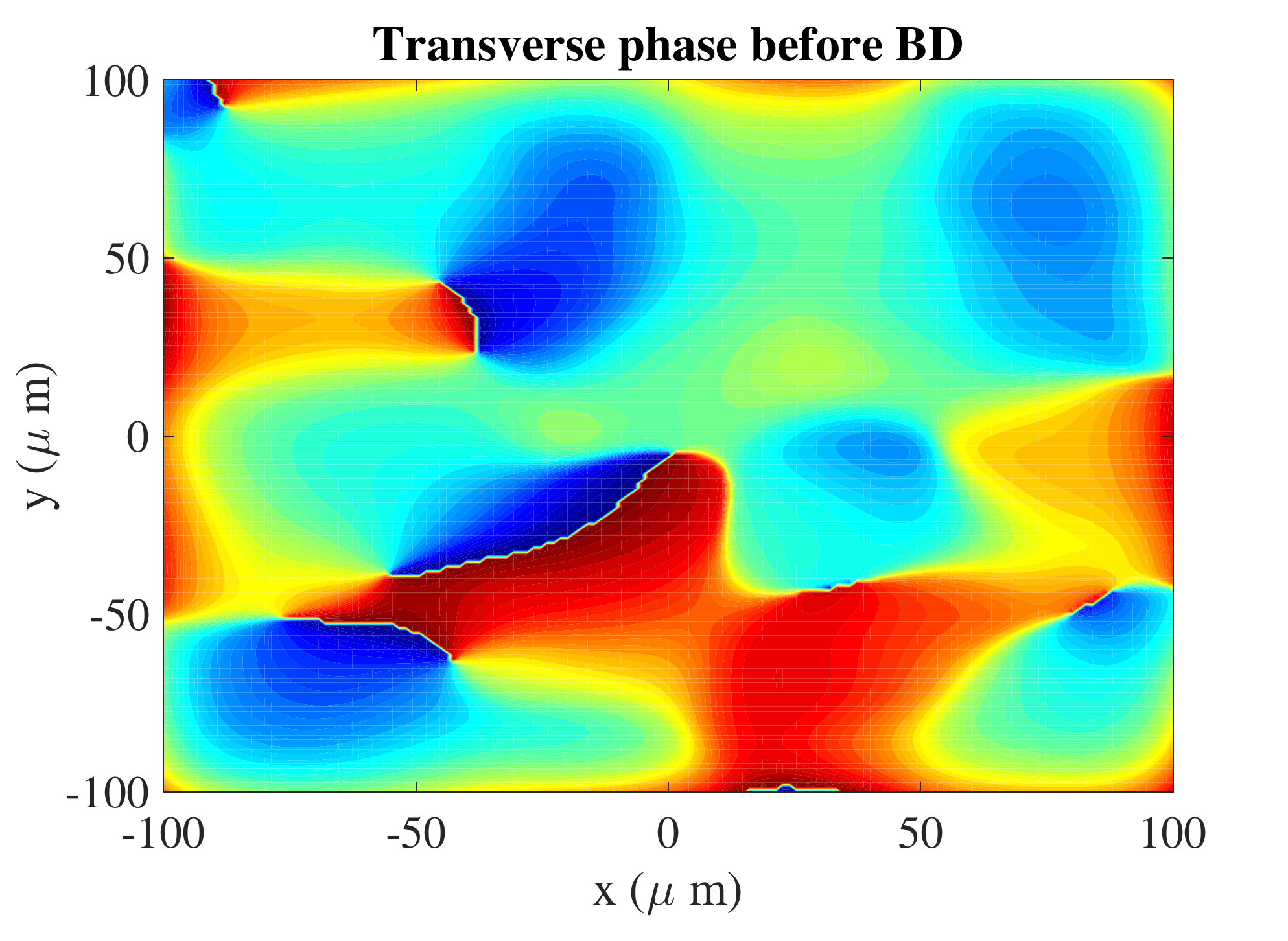}}
   \subfigure{\includegraphics*[width=100pt]{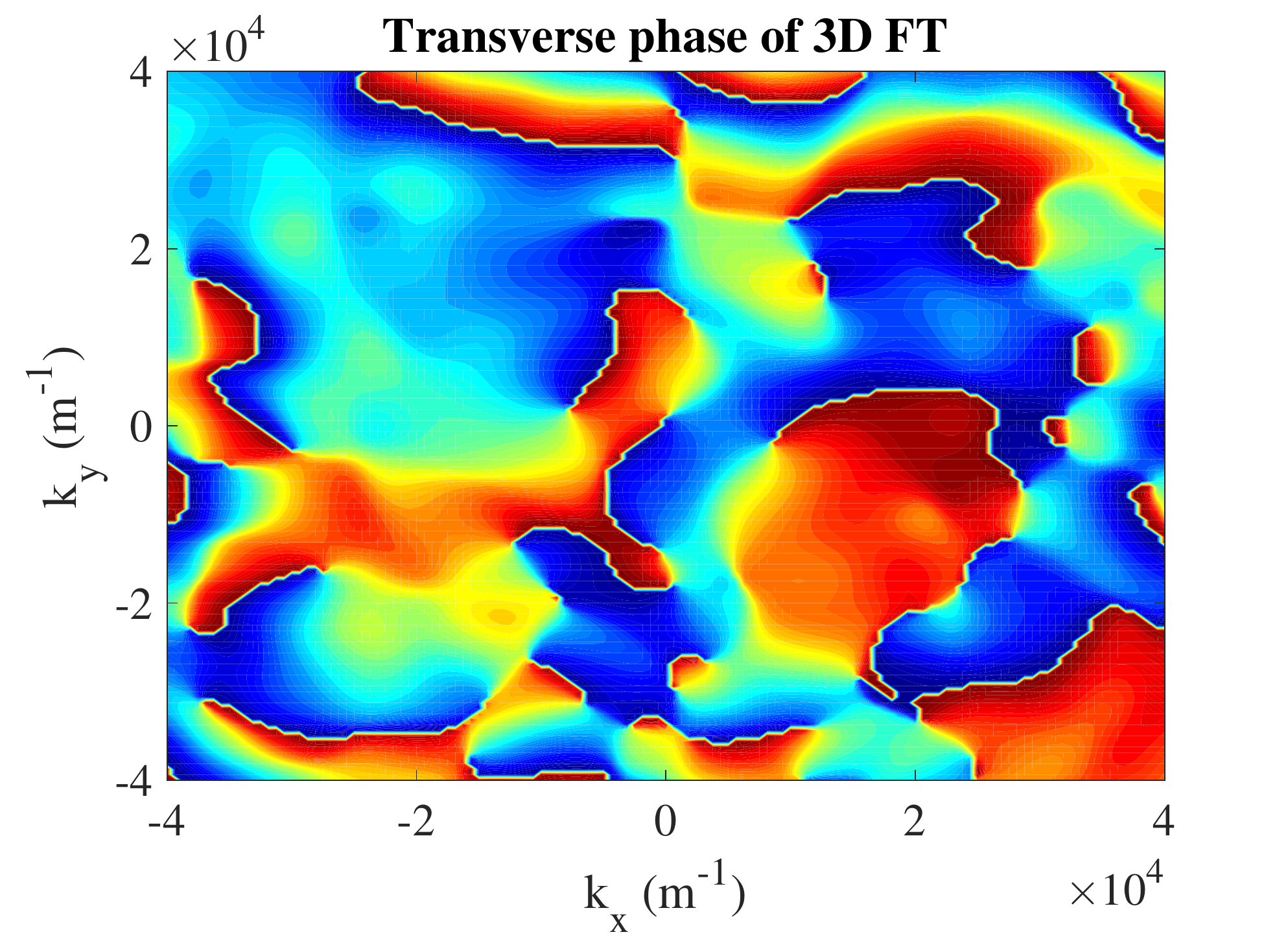}}
   \subfigure{\includegraphics*[width=100pt]{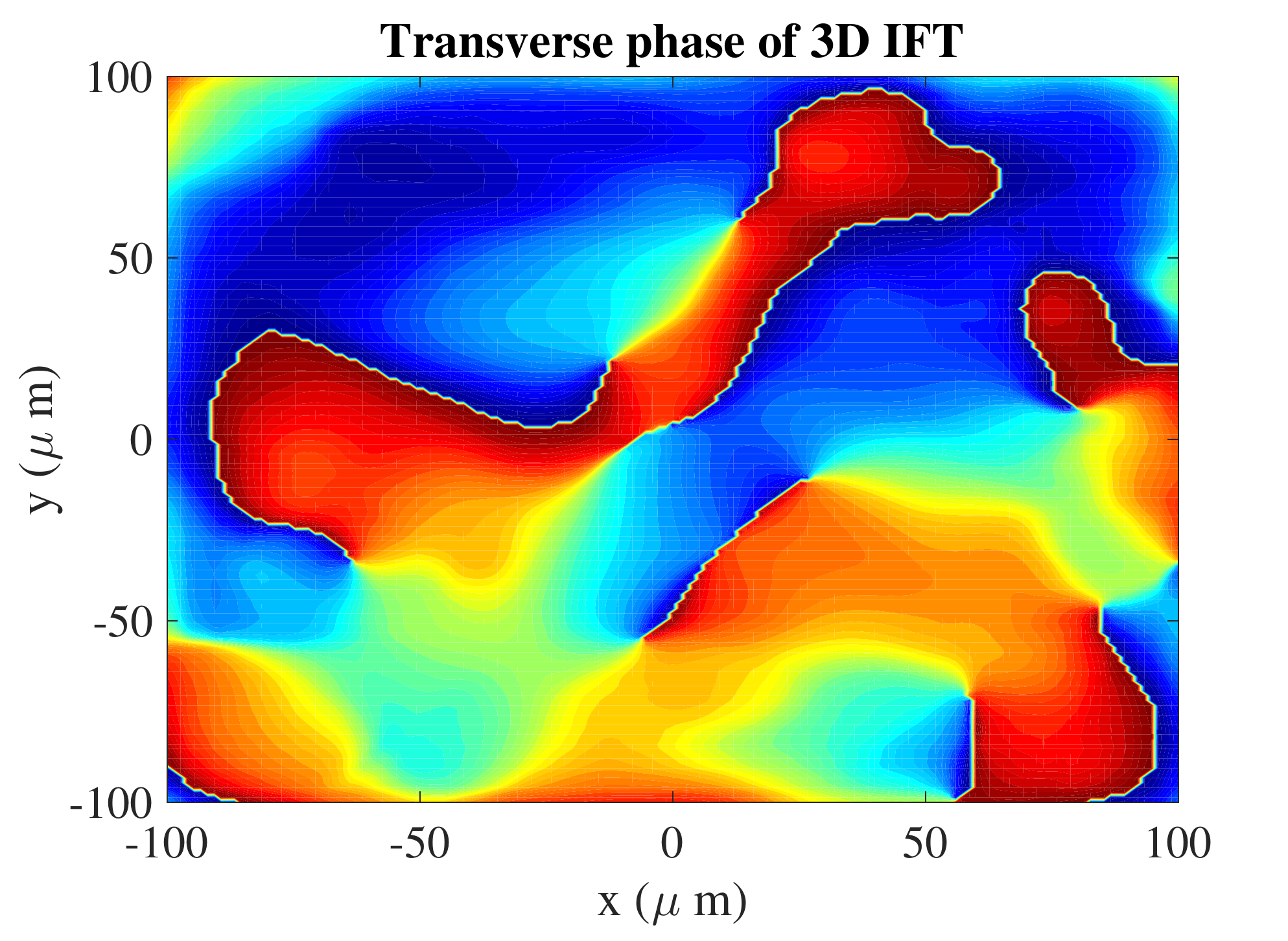}}
   \subfigure{\includegraphics*[width=100pt]{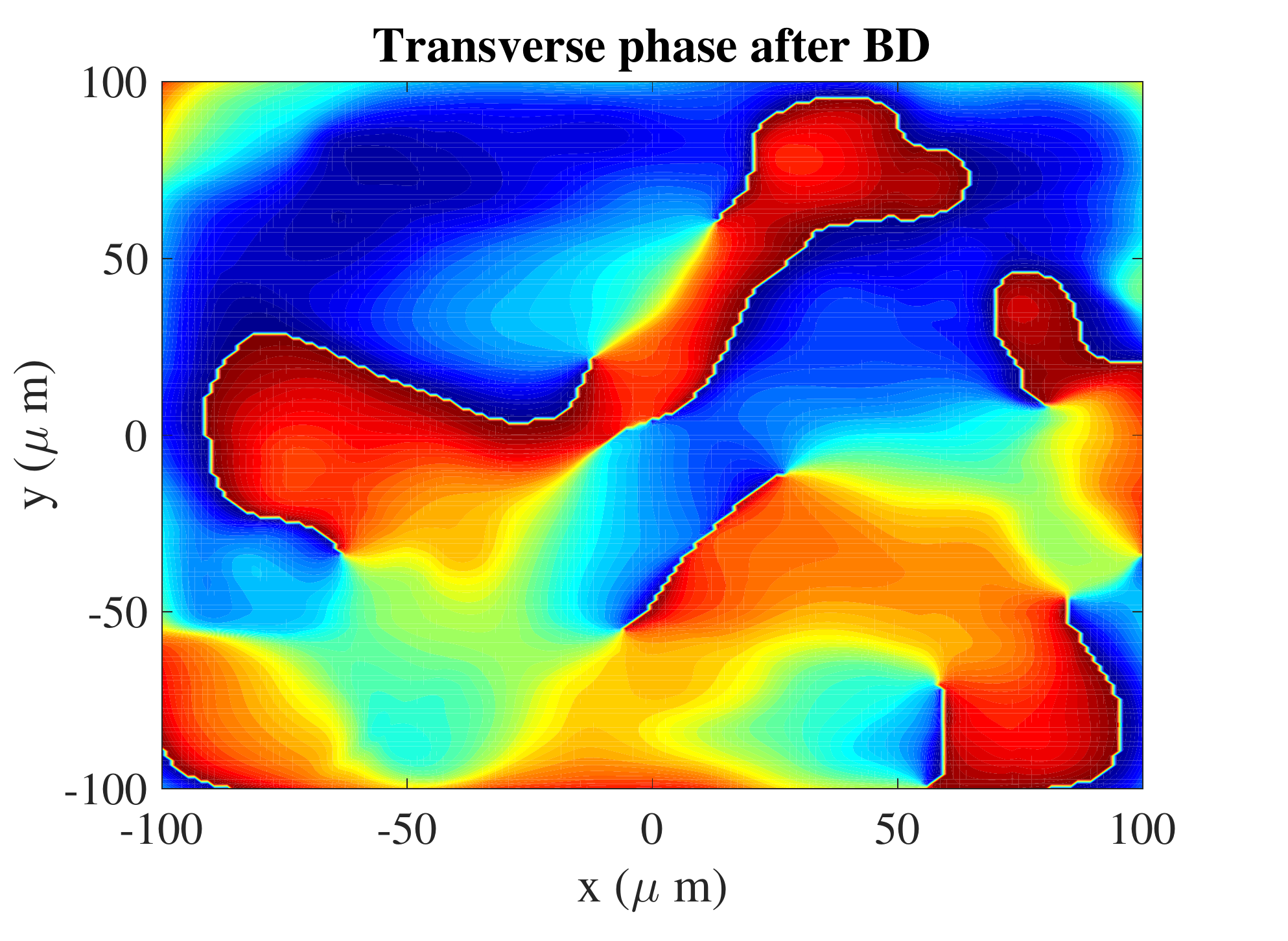}}
   \caption{The Bragg diffraction results of BRIGHT simulation for 1 kA XFELO at 10 passes. The first column shows the initial light pulse field longitudinal, transverse, and phase profile; the second column plots the corresponding figures after 3D Fourier transform; the third column shows the field distribution after Bragg diffraction by directly 3D approach (b); and the final column presents the results of new method (c).}
   \label{3d}
\end{figure*}

\begin{figure}[!htb]
   \centering
   \includegraphics*[width=220pt]{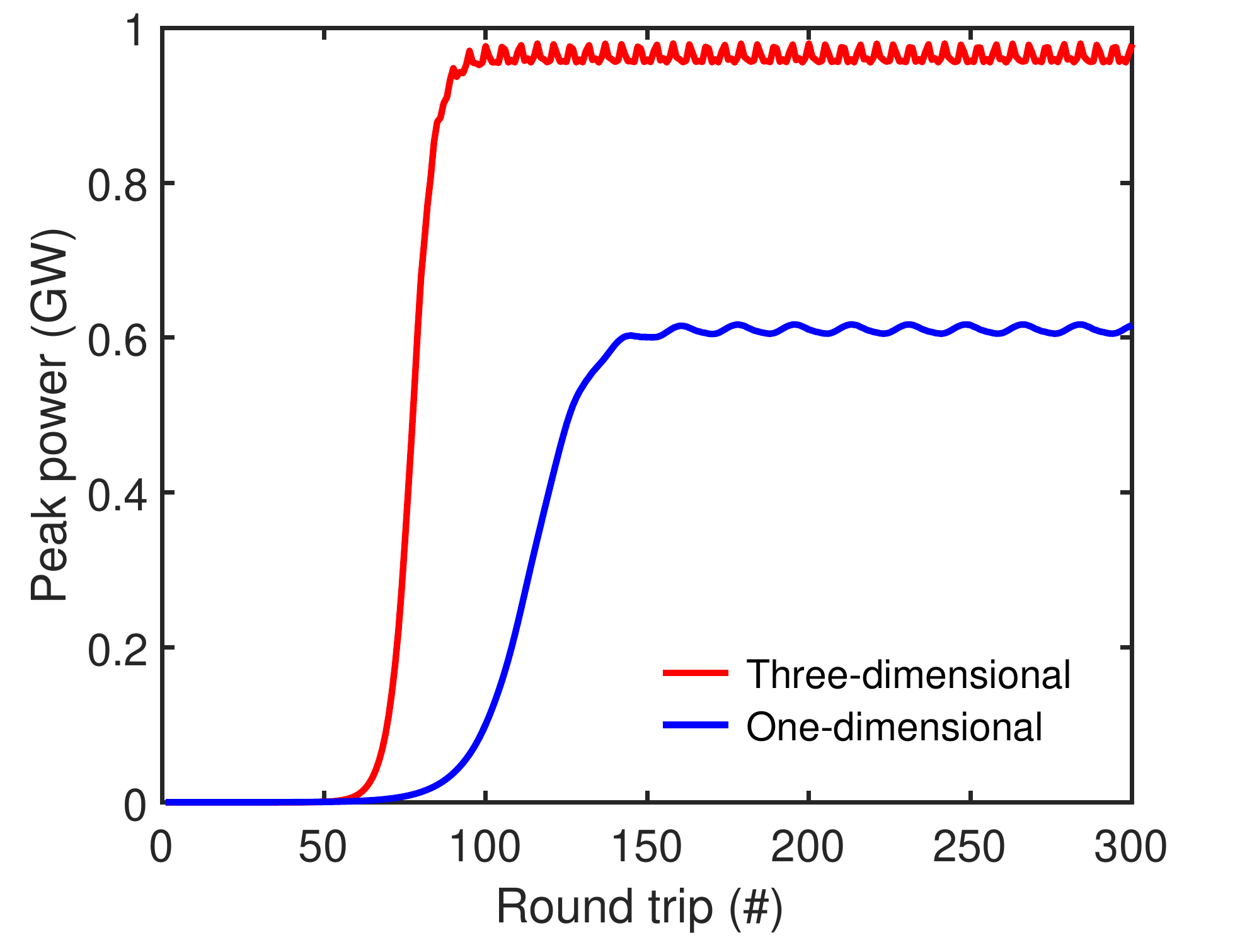}
   \caption{The comparison of one-dimension traditional simulation results with three-dimensional new method.}
   \label{compare}
\end{figure}

The BRIGHT results of scheme (c) is compared with (b) in Fig.~\ref{3d}: the first column represents the initial X-ray field distribution; the second column shows the corresponding 3D FT; the third column demonstrates the final radiation after Bragg diffraction of approach (b); while the final column is the results calculated by approach (c). The validity is confirmed by the similarity of the radiation power and phase distribution. Note that the small discrepancy is mainly due to the limited range we have covered in the frequency domain in scheme (b) which leads to the loss of fractional high frequency signals. Fig.~\ref{compare} shows the results of classical one-dimensional and BRIGHT for 1 kA high peak current mode XFELO simulation of SCLF, the reasonable agreement implies the validation of BRIGHT.  The three-dimensional simulation preserves transverse phase distribution of the radiation, and is benefit for eigenmode formation as expected. The simulation reveals that 3D simulation gives smaller radiation size and higher single pass gain, thus generates larger radiation power. For low current mode, the 3D Bragg diffraction effects proved to be indispensable, due to the much longer undulator implemented, the drift space is suppressed and the optical cavity is more sensitive. The conventional method undermines the resonator stable conditions and prevent the X-ray power from growing. Thus the 3D simulation is essential for low current XFELO simulation. With the discussions above, we expect that 3D XFELO simulations by BRIGHT would be prevalent. Note that the applications are not limited for XFELO, it is useful for simulations of other X-ray Bragg diffraction such as hard X-ray self-seeding FEL as well.

\section{\label{sec:level4}High current mode XFELO}
\begin{figure*}[!htb]
   \centering
   \subfigure{\includegraphics*[width=145pt]{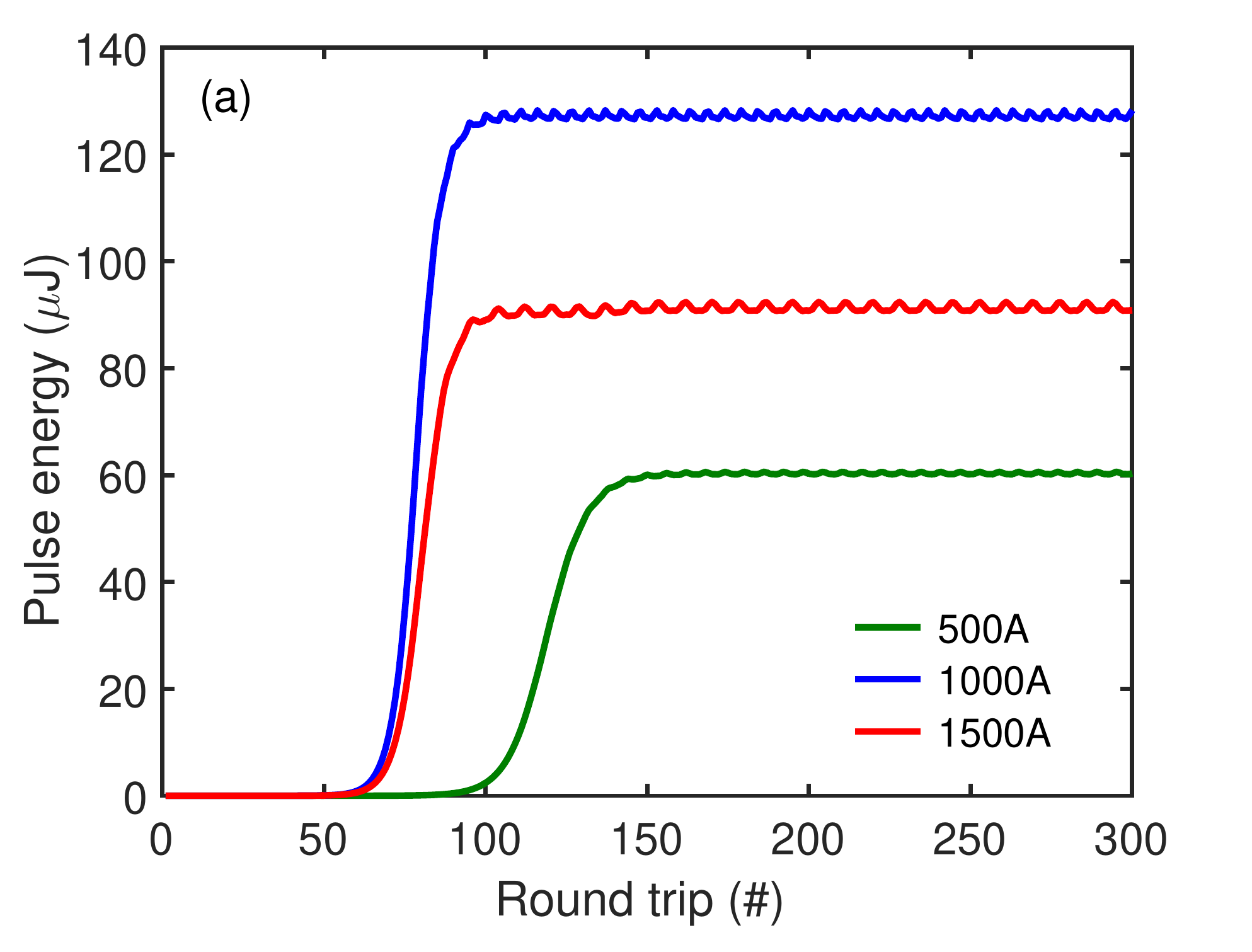}}
   \subfigure{\includegraphics*[width=145pt]{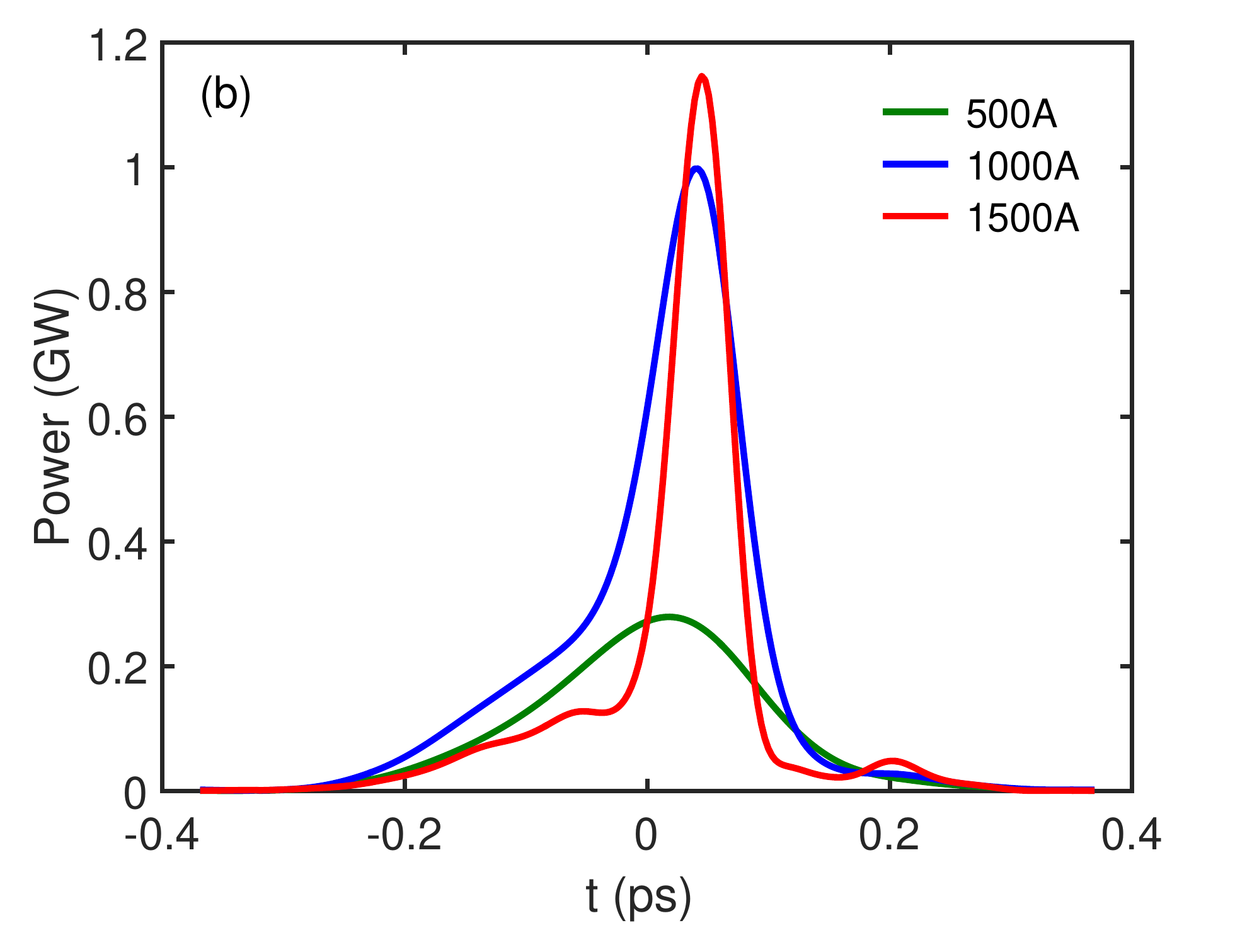}}
   \subfigure{\includegraphics*[width=145pt]{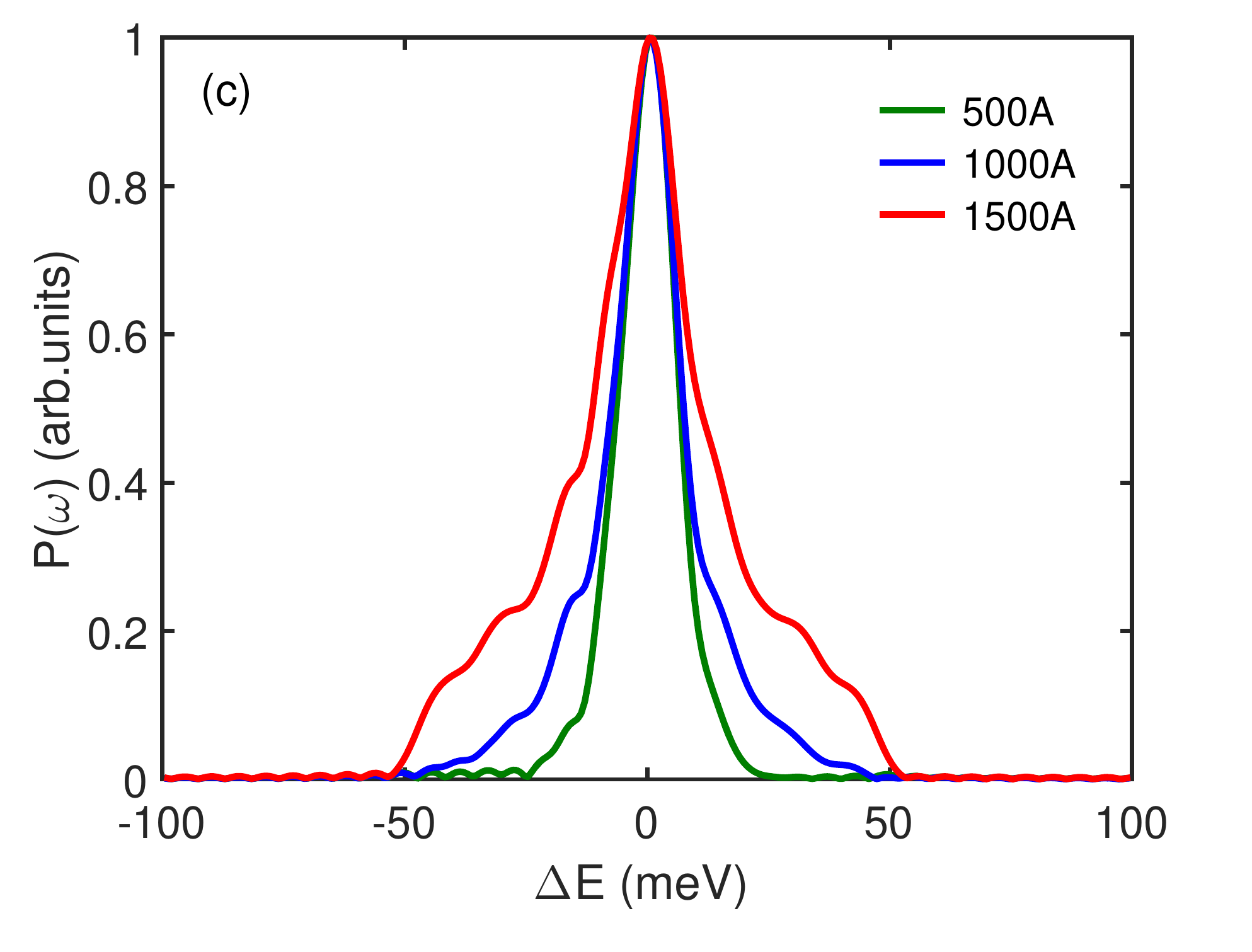}}
   \caption{The results of high peak current mode XFELO. The output energy growth (a), output power profile (b) and spectrum (c) for different electron beam peak current.}
   \label{high_energy}
\end{figure*}
The aforementioned analysis is utilized for the design of XFELO in SCLF. Candidate parameters are first calculated by a fast FEL oscillator code refer to \cite{li2017simplified}, in which the theoretical analysis produces approximately results rapidly. Then the parameters are simulated and optimized by GENESIS, OPC and BRIGHT. For high current modes, three undulator segments are employed, and sapphire mirror reflectivity is 95\% for the upstream and 80\% for the downstream with thickness equal to 70$\mu$m and 47 $\mu$m, respectively. The output energy as a function of round trip pass is demonstrated in Fig.~\ref{high_energy} (a). As expected, the power grows exponentially before saturation, and presents a stable X-ray output. The 1 kA XFELO generates largest pulse energy 128 $\mu J$, which is followed by 1500 A, and then the 500 A. This is due to that high peak current leads to short bunch duration, which broadens the corresponding spectrum bandwidth. The fractional photons energy beyond sapphire crystal Bragg reflection bandwidth cannot be accumulated. Thus the higher peak current is, the larger single pass gain is, yet more power is cut off during the interaction with mirrors. In addition, the normal peak current for SCLF is 1500 A with 100 pC charge, the results reveal that XFELO is able to operation with lower peak current. Furthermore, the energy fluctuation demonstrated in Fig.~\ref{high_energy}(a), might origins from the ``respiration'' of X-ray pulse length.

Fig.~\ref{high_energy} (b) and (c) show the power and spectrum of different peak current XFELO. For the 1 kA XFELO, the spectrum bandwidth is 15.6 meV (FWHM), which results from the joint efforts of crystal Bragg reflection bandwidth and electron beam duration. The X-ray pulse temporal length is around 0.09 ps, which gives a 0.35 time-bandwidth product, nearly the value of the Fourier transform limit (0.44) for a Gaussian pulse profile. In addition, with same configurations of electron beams and undulators, SASE generates 111 $\mu J$ X-ray pulse with peak brilliance $3.4\times10^{32}\,(photons/s/mm^2/mrad^2/0.1\% BW)$, while XFELO produces X-ray with peak brilliance $2.1\times10^{35}\,(photons/s/mm^2/mrad^2/0.1\% BW)$, which is three orders of magnitude higher than SASE.

\section{\label{sec:level5}Low current mode XFELO}
\begin{figure*}[!htb]
   \centering
   \subfigure{\includegraphics*[width=145pt]{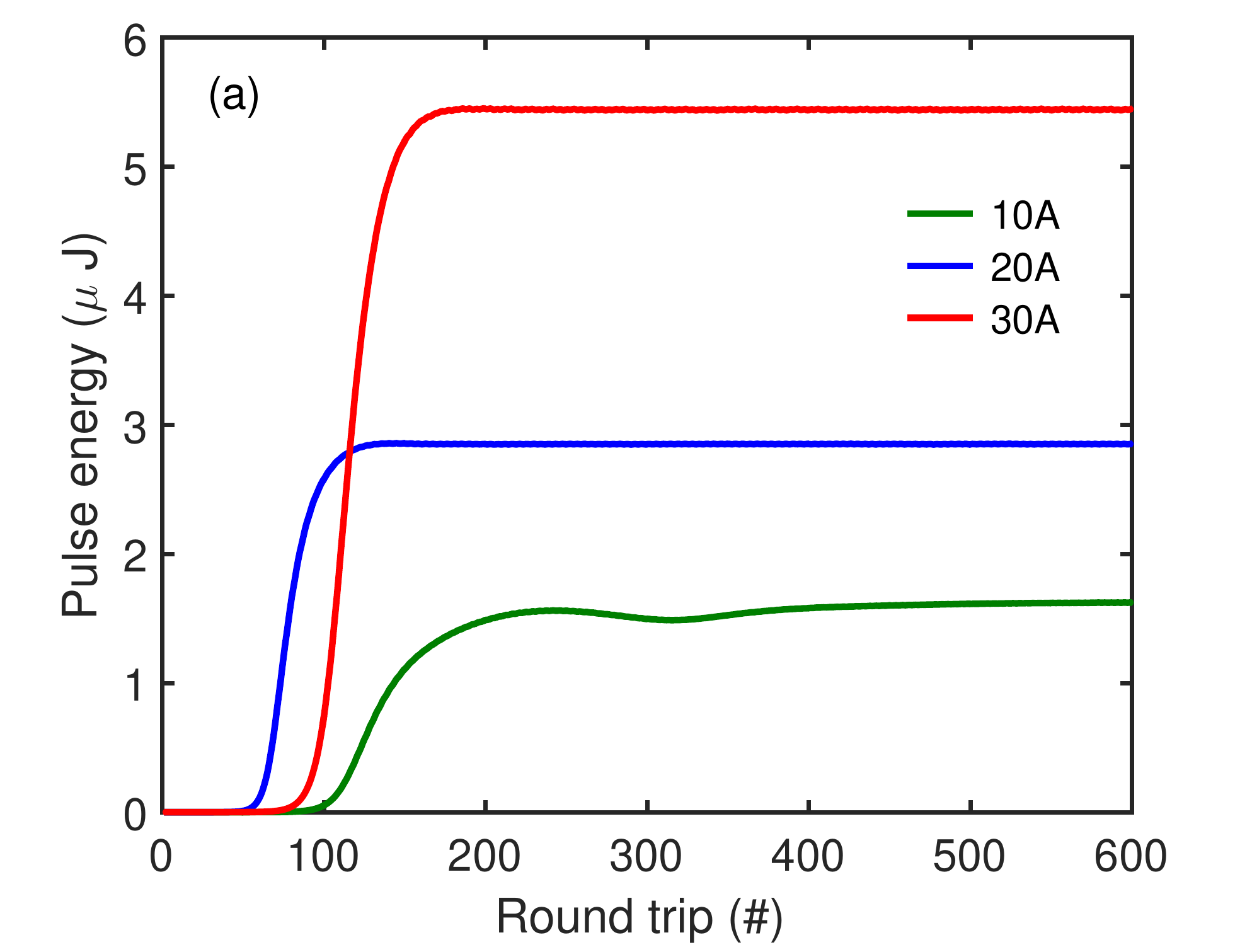}}
   \subfigure{\includegraphics*[width=145pt]{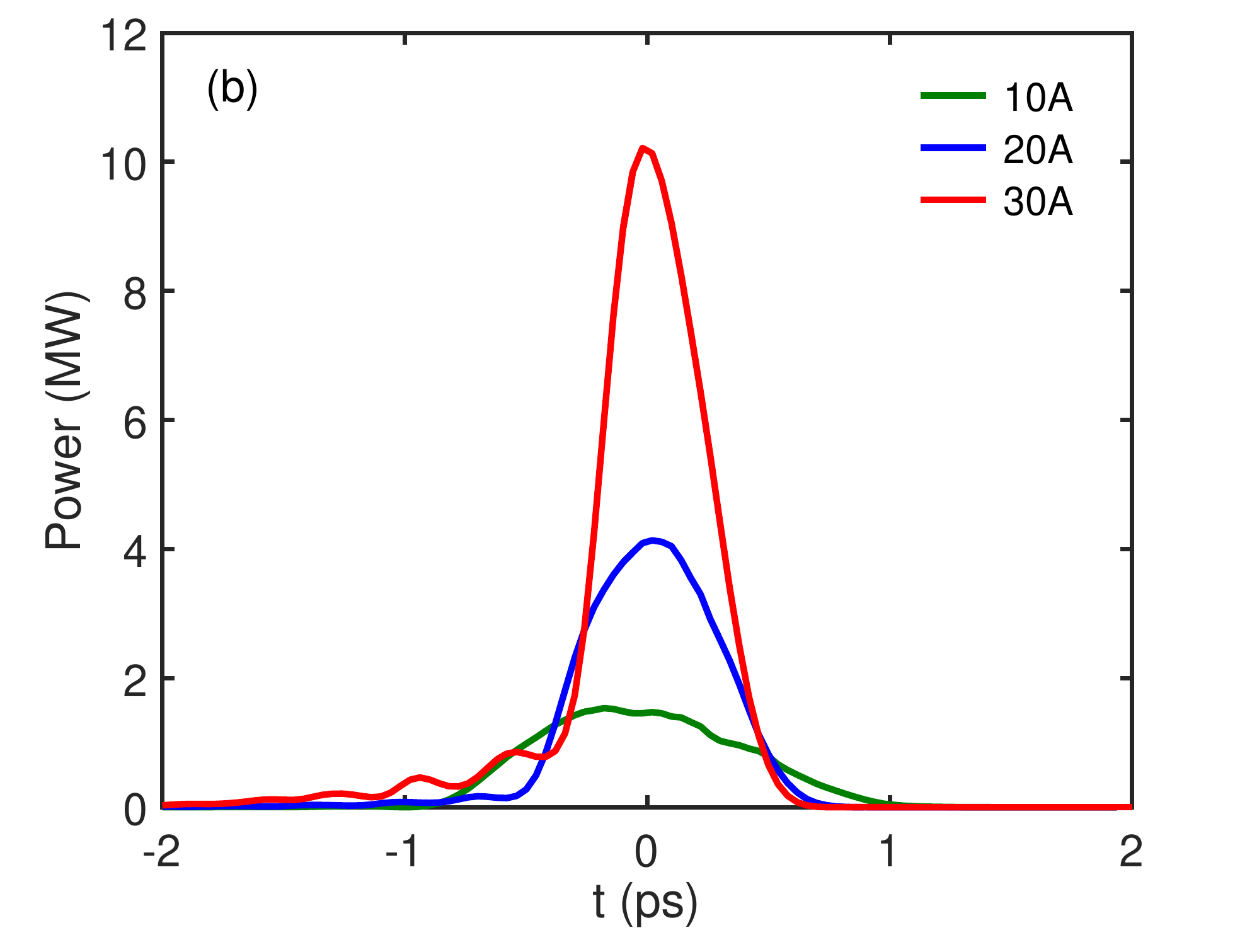}}
   \subfigure{\includegraphics*[width=145pt]{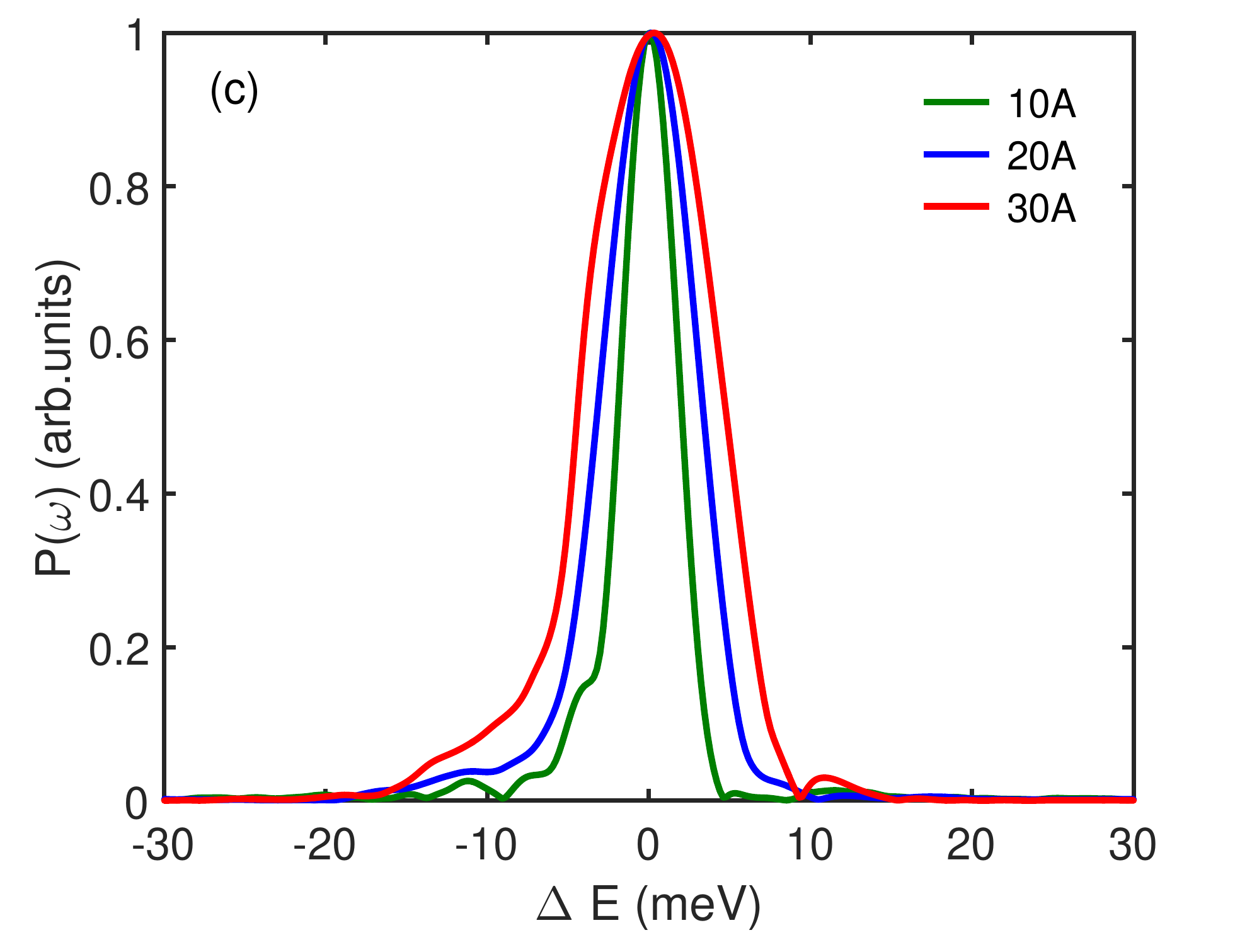}}
   \caption{The results of low peak current mode XFELO. The output energy growth (a), output power profile (b) and spectrum (c) for different electron beam peak current.}
   \label{low_energy}
\end{figure*}
The same methods are used to investigate the XFELO performance at 10, 20, 30 A peak current. To obtain enough single pass gain, the corresponding undulator segments demanded are 10, 8, 6. The optimum mirror reflectivity is nearly 93\% for downstream with thickness equals 65 $\mu$m and 95\% for upstream, with the crystal Darwin bandwidth nearly 13 meV. Fig.~\ref{low_energy} (a) shows the output light pulse energy evolves pass-by-pass of different peak current. It demonstrates that the pulse power increases exponentially before saturation. The maximum final output pulse energy is nearly 5.5 $\mu$J for 30 A. 

For the low current mode, electron beams charge in SCLF is 20 pC, which means the corresponding electron bunch length (FWHM) for different peak current are 2.00, 1.00, 0.67 ps, respectively. The output light pulse power profiles and spectrum are shown in Fig.~\ref{low_energy} (b) and (c). The pulse duration reduces as the peak current increases, the spectrum, however, does not change much, which indicates that the output X-ray is not at the Fourier transform limit. 
The output light pulse duration is nearly the same as the corresponding electron bunches length, while the spectrum bandwidth is about 9.4 meV, with peak brilliance nearly $3.6\times10^{33}\,(photons/s/mm^2/mrad^2/0.1\% BW)$ for 30 A XFELO.

\section{\label{sec:level7}Conclusion}
A systematical process for designing and optimization of XFELO is proposed. On the one hand, for optimization of transverse beam size, FODO lattice is used for average electron beam transverse size suppression; while the CRLs are employed to preserve the stability of optical cavity, whose transverse mode matches to electron bunch. On the other hand, suitable mirror reflectivity is chosen to couple out X-ray effectively and to feedback enough power to ensure the cavity pulse energy grows up. Meanwhile, the undulator should offer sufficient single pass gain to compensate the cavity loss.

SCLF is expected to be the first hard X-ray FEL facility in China. The quasi-CW high quality electron beam is appropriate for XFELO operation. XFELO examples based on SCLF is studied with high current mode ($\sim$ 1 kA) and low current mode ($\sim$ 10 A). We developed a novel simple three-dimensional Bragg diffraction codes BRIGHT, and 3D self-consistent numerical simulations are carried out by the combination of GENESIS, OPC and BRIGHT. The simulation results prove the design approaches to be feasible and efficient. According to our results, SCLF XFELO is able to generate 128 $\mu$J, 1 GW fully coherent stable X-ray with 1 kA electron beam peak current; and nearly several $\mu$J, 10 MW X-ray with a few dozens of A peak current. The peak brilliance of high current mode XFELO is nearly 3 orders of magnitude larger than SASE. It is worth to note that the design process is suitable for other FEL machines, and the results is useful for construction of XFELO test facility in SCLF as well as the facilities around the world. In addition, the parameters are useful for the investigation of other advanced proposals based on XFELO, such as cascaded single-pass XFELO \cite{deng2016proposals}. The following work would be to considering practical technical problems like the crystal thermal loading tolerance of X-ray interaction.

\section*{Acknowledgments}
The author would like to thank B.~Liu, D.~Wang and Z.~Zhao for helpful discussions on SCLF project; K. J.~Kim and W. L.~Qin for enthusiastic discussions on XFELO physics; and Y.~Shvyd'ko, R.~Lindberg for providing information and related parameters of crystal. This work was partially supported by the National Natural Science Foundation of China (11775293) and Ten Thousand Talent Program.


\section*{References}

\bibstyle{elsarticle-num}
\bibliography{mybibfile}

\end{document}